\documentclass[aps,twocolumn,showpacs,superscriptaddress,groupedaddress,nofootinbib]{revtex4}  % for review and submission
\usepackage{graphicx}  % needed for figures
\usepackage{dcolumn}   % needed for some tables
\usepackage{bm}        % for math
\usepackage{amssymb}   % for math
\usepackage{amsmath}   % for math
\usepackage{braket}
\usepackage{footnote}
\usepackage{subfigure}
\usepackage{color}
\usepackage{here}
\usepackage{hyperref}
\usepackage{multirow}
\usepackage{enumitem}

\usepackage[T1]{fontenc}
\usepackage{color}
\usepackage[normalem]{ulem}

\usepackage[utf8]{inputenc} 

\begin{document}

% The following information is for internal review, please remove them for submission
% the following line is for submission, including submission to the arXiv!!
%\hspace{5.2in} \mbox{Fermilab-Pub-04/xxx-E}

\title{Large $N_c$ scaling of meson masses and decay constants}

\date{\today}

%\author{A.~Donini}
%\affiliation{IFIC (CSIC-UVEG), Edificio Institutos Investigaci\'on, 
%Apt.\ 22085, E-46071 Valencia, Spain}
\author{P.~Hern\'andez}
\affiliation{IFIC (CSIC-UVEG), Edificio Institutos Investigaci\'on, 
Apt.\ 22085, E-46071 Valencia, Spain}
\author{C.~Pena}
\affiliation{Departamento de F\'{\i}sica Te\'orica and Instituto de F\'{\i}sica Te\'orica UAM-CSIC,
Universidad Aut\'onoma de Madrid, E-28049 Madrid, Spain}
\author{F.~Romero-L\'opez}
\affiliation{IFIC (CSIC-UVEG), Edificio Institutos Investigaci\'on, 
Apt.\ 22085, E-46071 Valencia, Spain}

%\preprint{IFIC/16-39}
%\preprint{IFT-UAM/CSIC-16-063}
%\preprint{FTUAM-16-26}

\begin{abstract}
We perform an  \textit{ab initio} calculation of the  $N_c$ scaling of the low-energy couplings of the chiral Lagrangian of low-energy strong interactions, extracted from the mass dependence of meson masses and decay constants. We compute these observables on the lattice with four degenerate fermions, $N_f=4$, and varying number of 
colours, $N_c=3-6$, at a lattice spacing of $a\simeq 0.075$ fm.  We find good agreement with the expected $N_c$ scaling and measure the coefficients of the leading and subleading terms in the large $N_c$ expansion. From the subleading $N_c$ corrections, we can also infer the $N_f$ dependence, that we use to extract 
 the value of the low-energy couplings for different values of
$N_f$. We find agreement with previous determinations at $N_c=3$ and $N_f=2, 3$ and also, our results support a strong paramagnetic suppression of the chiral condensate in moving 
from $N_f=2$ to $N_f=3$. 
\end{abstract}
\maketitle

%\section{\label{sec:level1}First-level heading}
% sections are not used for PRL papers

The 't Hooft limit of QCD \cite{tHooft:1973alw} is well known to capture correctly most of its non-perturbative features, such as confinement and chiral symmetry breaking. Large $N_c$ inspired approximations are often employed in  phenomenological approaches to hadron physics \cite{Bardeen:1986vp, Bardeen:1986uz,Chivukula:1986du,Sharpe:1987cx,Pich:1995qp,Hambye:2003cy,Buras:2014maa,Gisbert:2018tuf,Pelaez:2010er, Nebreda:2011cp}, but systematic errors from subleading $N_c$ corrections are only naively estimated. 

Lattice Field Theory offers the possibility of \textit{ab initio} explorations of the large $N_c$ limit of QCD, by simulating at different values of $N_c$ \cite{Lucini:2001ej,Bali:2013kia}. Several studies have already been performed. In Ref. \cite{Bali:2013kia} a thorough study of mesonic two-point functions was carried out in the quenched approximation, a limit that captures correctly the leading order terms in $N_c$, but modifies subleading corrections in an uncontrolled way. Furthermore, in Ref. \cite{DeGrand:2016pur} a similar study was performed for $N_c= 2-5$ using $N_f=2$ dynamical fermions at rather high pion masses.

In addition to the standard approach, the study of QCD in the large $N_c$ limit can also be achieved using reduced models (see \cite{Lucini:2012gg} for a review). In this context, there has been significant progress regarding the properties of mesons \cite{Perez:2018afi, Perez:2016qcv, Gonzalez-Arroyo:2015bya,Hietanen:2009tu,Narayanan:2005gh}. 

Besides, lattice simulations have been used to perform studies of various observables in theories with different number of colours, flavours or fermion representations in the context of Beyond-the-Standard-Model theories. Some recent results can be found in  \cite{Appelquist:2014jch,DeGrand:2015lna, Aoki:2016wnc, Appelquist:2016viq,Hansen:2017mrt,Nogradi:2019iek} and for recent reviews see \cite{Brower:2019oor, DelDebbio:2018szp}. 

In this work, we use previously generated lattice configurations with $N_c=3-6$ and four dynamical fermions. Our particular choice of $N_f$ has also advantages for weak matrix elements \cite{Donini:2016lwz}. On these ensembles, we compute meson masses and decay constants as a function of the quark mass at the different values of $N_c$. We fit these to chiral perturbation theory (ChPT) in order to extract the leading order and next-to-leading order low-energy chiral couplings (LECs). We then study their $N_c$ scaling and extract the first two terms in the 't Hooft series. Our study  builds on previous lattice determinations of the LECs for $N_c=3$ \cite{Dowdall:2013rya, Frezzotti:2008dr,Baron:2009wt,Brandt:2013dua, Gulpers:2015bba,Bazavov:2010hj,Beane:2011zm,Borsanyi:2012zv,Durr:2013goa,Boyle:2015exm,Baron:2011sf, Aoki:2008sm,Noaki:2008iy,Baron:2010bv,Drach:2017btk}, whose main results are summarized in \cite{Aoki:2019cca}.

 Interestingly, within the large $N_c$ expansion,  the $1/N_c$ corrections have a well-defined  linear dependence on $N_f$, while the 't Hooft limit is independent on $N_f$. Using this fact,  we can predict the low-energy couplings at different values of $N_f$ up to higher orders in $N_c$. This allows us to compare with previous determinations, and check the prediction of {\it paramagnetic} suppression at large $N_f$ of Refs. \cite{DescotesGenon:1999uh,Bernard:2012fw}.

This paper is organized as follows. First, we describe chiral perturbation theory predictions and the relation to the large $N_c$ limit in Section~\ref{sec:chpt}. In Section \ref{sec:setup},  we present the lattice setup that involves a mixed-action formulation. Next, we explain our scale setting procedure at different $N_c$ consistent with 't Hooft scaling in Section~\ref{sec:scale}. In Section~\ref{sec:fits} we present the results of our chiral fits to the meson mass and decay constant, first at fixed $N_c$ and then combined with the large $N_c$ expansion. We also present {results} for theories with different values of $N_f$, compare with previous literature and discuss systematic uncertainties. We conclude in Section~\ref{sec:conclu}.

\section{Chiral perturbation Theory predictions}
\label{sec:chpt}

The light spectrum of QCD is the result of the pattern of spontaneous chiral symmetry breaking, $SU(N_f)_L \times SU(N_f)_R \rightarrow SU(N_f)_{L+R}$. ChPT represents accurately the dynamics of the expected pseudo-Nambu-Goldstone bosons (pNGB), i.e., the lightest non-singlet multiplet of pseudoscalar mesons (the octet for $N_f=3$),  at sufficiently small quark masses.
The increase in the number of colours while keeping the 't Hooft coupling constant, $\lambda = g^2 N_c$, is not expected to modify these features. On the other hand, in the large $N_c$ limit, QCD reduces to a theory of narrow and non-interacting resonances and, as a result, the interactions of pNGB within the effective theory decrease with $N_c$, improving the convergence of the perturbative series.
One complication of the large $N_c$ expansion is the role of the singlet pseudoscalar meson, i.e., the $\eta'$. Its mass originates in the explicit $U(1)_A$ breaking by the anomaly. In QCD this contribution to the mass is at the cutoff scale of the chiral effective theory and it is therefore integrated out. However, the anomalous contribution to the singlet mass decreases with $N_c$ and  in the 
large $N_c$ limit the $\eta'$ becomes degenerate with the remaining pNGBs. The effective theory should consequently include an additional singlet pseudoscalar meson in the spectrum. 
The corresponding effective theory has been studied long ago \cite{DiVecchia:1980yfw,Rosenzweig:1979ay,Witten:1980sp,Kawarabayashi:1980dp,HerreraSiklody:1996pm,Kaiser:2000gs}. A new power-counting is needed which involves a simultaneous expansion in $1/N_c$ and the usual chiral expansion in the quark mass and momenta. A consistent power counting was implemented in refs.  \cite{HerreraSiklody:1996pm,Kaiser:2000gs}:
\begin{eqnarray}
{\mathcal O}(\delta) \sim {\mathcal O}(p^2) \sim {\mathcal O}(m_q)\sim {\mathcal O}(m_\pi^2) \sim {\mathcal O}(N_c^{-1}).
\end{eqnarray}
In the following we will concentrate on the non-singlet multiplet masses and decay constants. We now compare the usual $SU(N_f)$ ChPT to the $U(N_f)$ ChPT for these observables.  

\subsection{$SU(N_f)$ effective theory}
At Leading Order (LO) in the standard $SU(N_f)$ chiral expansion there are only two couplings for any number of degenerate flavours, related to the chiral condensate and the meson decay constant. At Next-to-Leading Order (NLO), and for an arbitrary number of degenerate flavours ($N_f>3$), 13 more LECs are needed, but only two combinations enter in the observables of interest. For $N_f$ degenerate flavours, ChPT predicts at NLO \cite{Bijnens1,Bijnens2,Bijnens3}:
\begin{align}
\begin{split}
F_\pi = {F} \Bigg[&1   - {\frac{N_f}{2} \frac{M_\pi^2}{(4 \pi F_\pi)^2}\log \frac{M_\pi^2}{\mu^2}}     \\ &+4 \frac{M_\pi^2}{F_\pi^2} \Big( L_5^r  +  N_f L_4^r  \Big)  \Bigg], 
\label{eq:Fpi1}
\\
\end{split}
\end{align}
and 
\begin{align}
\begin{split}
M^2_\pi &= 2B m \Bigg[1  + {\frac{1}{N_f} \frac{M_\pi^2}{(4 \pi F_\pi)^2}\log \frac{M_\pi^2}{\mu^2}} \\ & +8\frac{M_\pi^2}{F_\pi^2} \Big(  2L_8^r - L_5^r + N_f (2 L_6^r  - L_4^r )   \Big)  \Bigg], \label{eq:Mpi1}
\end{split}
\end{align}
in terms of the LO couplings, $B, F$, and the NLO Gasser-Leutwyler coefficients,  $L^r_{4,5,6,8}(\mu)$, defined at the renormalization scale $\mu$. 

Eqs.~(\ref{eq:Fpi1}-\ref{eq:Mpi1}) are valid for an arbitrary number of colours, but the LECs scale with $N_c$ as (for a review see \cite{Manohar}):
\begin{eqnarray}
\begin{array}{ll}
O(N_c): F^2, L_5, L_8;& 
 O(1): B, L_4, L_6.
 \end{array}
 \label{eq:nc}
\end{eqnarray}
Loop corrections are suppressed in $1/F_\pi^2 = O(1/N_c)$, and hence the loop expansion is expected to converge better at larger $N_c$. 

Keeping only leading and subleading dependence on $N_c$ a convenient parametrization is
\begin{align}
F = \sqrt{N_c} \left( {F}_0 + \frac{{F_1}}{N_c}  \right), \ \ \ \ \ B =   {B}_0 + \frac{{B_1}}{N_c}, 
\label{eq:lonc}
\end{align}
and
\begin{eqnarray}
& L_5  + N_f L_4 \equiv L_F = N_c L_F^{(0)}  + L_F^{(1)}, \\
 & 2 L_8 - L_5 + N_f (2 L_6 - L_4 )\equiv L_M = N_c L_M^{(0)}  + L_M^{(1)}.
\label{eq:lecsnc}
\end{eqnarray}
Note that according to the scaling of Eq.~(\ref{eq:nc}) and the definition of Eq. \ref{eq:lecsnc}:
\begin{align}
\begin{split}
&L_F^{(0)} = {L_5 \over N_c} + {\mathcal O}\left({1 \over N_c}\right), \\ \ \ &L_M^{(0)} = {2 L_8 - L_5 \over N_c} + {\mathcal O}\left({1 \over N_c}\right).
\end{split}
\end{align}
The NNLO Lagrangian of the $SU(N_f)$ theory is also known \cite{Bijnens1,Bijnens2,Bijnens3}. At this order we will instead only use the $U(N_f)$, to which we now turn.

\subsection{$U(N_f)$ effective theory}
In the $U(N_f)$ ChPT at NLO, i.e., ${\mathcal O}(\delta^1)$, the result can be read from Eqs.~(\ref{eq:Fpi1}) and (\ref{eq:Mpi1}) and the different $N_c$ scalings of the LECs in Eqs.~(\ref{eq:lonc}) and (\ref{eq:lecsnc}):
\begin{align}
\begin{split}
F_\pi = \sqrt{N_c} \left(F_0+{F_1\over N_c}\right) &\Bigg[1    + 4\frac{M_\pi^2}{F_\pi^2}  N_c { L}^{(0)}_F  +{\mathcal O}(\delta^2)  \Bigg],
\label{eq:Fpileut}
\\
\end{split}
\end{align}
and 
\begin{align}
\begin{split} \hspace{-0.1cm}
M^2_\pi = 2 \left(B_0+ {B_1\over N_c}\right) m \Bigg[1   +  8\frac{M_\pi^2}{F_\pi^2}  N_c {L}_M^{(0)} +{\mathcal O}(\delta^2) \Bigg]. \label{eq:Mpileut}
\end{split}
\end{align}

The NLO corrections are not enough to explain the data in this case, therefore going to NNLO is essential.  At NNLO new features appear, because the singlet contributes to the mass loop corrections. The necessary results can be found in Ref.~\cite{Guo:2015xva}. For degenerate flavours, they simplify to:
\begin{eqnarray}
&  F_\pi &=\sqrt{N_c} \left( {F}_0+ {{F}_1\over N_c} + {{ F}_2 \over N_c^2} \right) \Bigg[1    - {\frac{N_f}{2} \frac{M_\pi^2}{(4 \pi F_\pi)^2}\log \frac{M_\pi^2}{\mu^2}} \nonumber \\
&+&4 \frac{M_\pi^2}{F_\pi^2} \Big( N_c {L}_F^{(0)} + {L}_F^{(1)} \Big) + N_c^2 K_F^{(0)} \left({M_\pi^2\over F_\pi^2}\right)^2 \\&+& \ {\mathcal O}(\delta^3)  \Bigg] , \nonumber
\label{eq:Fpi}
\end{eqnarray}
and 
\begin{eqnarray}
&M^2_\pi &= 2  m \left( { B}_0+ {{ B}_1\over N_c} + {{B}_2 \over N_c^2} \right)  \Bigg[ \nonumber  \\
&1&+ {\frac{1}{N_f} \frac{M_\pi^2}{(4 \pi F_\pi)^2}\log \frac{M_\pi^2}{\mu^2}} - {\frac{1}{N_f} \frac{M_{\eta'}^2}{(4 \pi F_\pi)^2}\log \frac{M_{\eta'}^2}{\mu^2}}\\ & +&8\frac{M_\pi^2}{F_\pi^2} \Big( N_c {L}_M^{(0)} + {L}_M^{(1)} \Big) + N_c^2 K_M^{(0)} \left({M_\pi^2\over F_\pi^2}\right)^2 +{\mathcal O}(\delta^3) \Bigg] \nonumber  ,\label{eq:Mpi}
\end{eqnarray}
where  $K_{F,M}^{(0)}$ are combinations of  $L_F^{(0)}$, $L_M^{(0)}$ and new LECs that appear in the $U(N_f)$ case. For details see \cite{Guo:2015xva}. Note that for degenerate
quarks, there is no $\eta$-$\eta'$ mixing. 

The $\eta'$ mass in this expression can be taken in the large $N_c$ limit, where it is given by the Witten-Veneziano formula:
\begin{eqnarray}
M_{\eta'}^2 = M_\pi^2 + {2 N_f \over F^2} \chi_t \equiv M_\pi^2 + M_0^2, \label{eq:WV}
\end{eqnarray}
where $\chi_t$ is the topological susceptibility in pure Yang-Mills, recently computed in the large $N_c$ limit in Ref.~\cite{Ce:2016awn}. 

Note that even though we use the same notation for the LECs in both chiral expansions, they are different: in the $SU(N_f)$ ChPT the LECs encode the effects of integrating out the $\eta'$. The matching of the two theories starts at NNLO \cite{Kaiser:2000gs,HerreraSiklody:1998cr} and only affects the coupling $B$ and $L_M^{(1)}$ of the above \cite{HerreraSiklody:1998cr,Kaiser:2000gs}:
\begin{align}
\begin{split}
\left[B\right]_{SU(N_f)} &= \left[B\right]_{U(N_f)} \left( 1- \frac{1}{N_f} \frac{M_0^2}{(4 \pi F_\pi)^2}\lambda_0 \right),\\
\left[L_M^{(1)}\right]_{SU(N_f)}  &= \left[L_M^{(1)}\right]_{U(N_f)} -{1 \over 8 N_f (4 \pi)^2}\left(\lambda_0+1\right),
\label{eq:match}
\end{split}
\end{align}
with $\lambda_0 = \log \frac{M_0^2}{\mu^2}$.

%In addition, one expects that the decay constant in the chiral limit is lower for $N_f=4$ than standard determinations with $N_f=3$ active flavours. This fact can be seen by treating the decoupling of the charm can be treated in ChPT \cite{Giusti:2004an}. This way, the decay constant in the chiral limit, $\bar{F}$ with $N_f=3$ is
%\begin{align}
%\begin{split}
%\frac{\bar{F}}{F} =   1- \frac{1}{2} \frac{\chi_{cc}}{(4 \pi F_0)^2} \bigg( \log \frac{\chi_{cc}}{2 {\mu}^2} -1 -(16 \pi)^2 L_4 \bigg) , \label{eq:charmdec}
%\end{split}
%\end{align}
%where $\chi_{cc} = 2 B m_c$. Of course, this can only be trusted in the region of validity of ChPT, but it serves as a  qualitative interpretation.

\subsection{$N_f$ versus $N_c$ dependence}

A diagrammatic analysis of fermion bilinear two point functions shows that within the large $N_c$ expansion, the leading order $N_c \rightarrow \infty$ limit is $N_f$ independent
and the NLO  is ${\mathcal O}(N_f/N_c)$. We should confirm this expectation also in ChPT formulae above, in particular given the explicit dependence on $N_f$. It turns out that within the $U(N_f)$ expansion, the large $N_c$ expansion yields the expected behaviour: the terms in $1/N_f$ exactly cancel when the large $N_c$ expansion is taken at fixed $M_\pi$. 
We expect therefore that the LECs should also satisfy this same scaling.

On the other hand within the $SU(N_f)$ expansion or in the $U(N_f)$ when $M_\pi \ll M_{\eta'}$, that is when the chiral limit is taken first, anomalous $1/N_f$ terms appear coming from 
an expansion in $M_\pi/M_{\eta'}$.  In the $U(N_f)$ expansion such dependence is explicit, but in the $SU(N_f)$ it permeates to the LECs which can no longer be assumed to have the expected ${\mathcal O}(N_f/N_c)$ dependence, as can be explicitly seen in the matching of $L_M^{(1)}$ in Eq.~(\ref{eq:match}).

This way, at the order we are working, we can assume the expected scaling in $N_f$ of the $U(N_f)$ and $SU(N_f)$ couplings except in the case of $[L_M^{(1)}]_{SU(N_f)}$. 
\section{Lattice setup \label{sec:setup}}

 We have generated ensembles for $SU(N_c)$ gauge theory with $N_f=4$ degenerate dynamical fermions, varying $N_c=3$-$6$, using the HiRep code \cite{hirep}. Some of them have been already presented in Ref.~\cite{Romero-Lopez:2018rzy}. We have chosen the Iwasaki gauge action (following previous experience with 2+1+1 simulations \cite{Finkenrath:2017cau,Alexandrou:2018egz}) and $O(a)$-improved\footnote{{For $N_c=3$, we take the perturbative value  of $c_{sw} = 1 + c^{(1)}_{sw} g^2$ from Ref. \cite{Aoki:2003sj}, where we use the plaquette-boosted coupling $g^2 = 2 N_c/(\beta P) = O(1/N_c)$. For other values of $N_c$, we use the fact that the one loop coefficient is dominated by the tadpole contribution, which is of order $N_c$ (see Eq. 58 in Ref. \cite{Aoki:2003sj}). This way, $c_{sw}$ is constant up to subleading corrections in $N_c$, which have an effect of $O(a^2/N_c)$ in physical observables. The full result cannot be easily reconstructed from Ref. \cite{Aoki:2003sj}.   }} Wilson fermions for the sea quarks. {Our simulations use the standard Hybrid Montecarlo (HMC) algorithm with Hasenbusch acceleration. We include five layers in each of the fermionic monomials. Interestingly, we observe that the tuning of the integrator at $N_c=3$ yields similar results at other values of $N_c$ (at similar pion mass) for the acceptance rate, which we keep at $80-90\%$. The computational cost of each step in Montecarlo time scales as $\sim N_c^2$, with the advantage of a more efficient parallelization at large $N_c$. }
 
In order to achieve automatic $O(a)$ improvement and avoid the need of a non-perturbative determination of normalization factors, we employ  maximally twisted valence quarks, i.e., the mixed-action setup  \cite{Bar:2002nr}  previously  used in Refs. \cite{Herdoiza:2017bcc,Bussone:2018ljj, Bussone:2018wki}. Maximal twist is ensured by tuning the untwisted bare valence mass, $m^{\rm v}$ to the critical value for which the valence PCAC mass is zero:
\begin{equation}
\lim_{m^{\rm v} \rightarrow m_{\rm cr}} m^{\rm v}_{\rm pcac} \equiv \lim_{m^{\rm v} \rightarrow m_{\rm cr}} \frac{\partial_0\braket{A^0(x) P(y)}}{2 \braket{P(x) P(y)}}= 0,
\end{equation}
where $A^\mu(x) \equiv \bar{\Psi}(x) \gamma^\mu \gamma^5 \Psi(x)$ and $P(x) = \bar{\Psi}(x) \gamma^5 \Psi(x)$. 
The bare twisted mass parameter $\mu_0$ is tuned such that the pion mass in the sea and valence sectors coincide, $M_\pi^{\rm v}= M_\pi^s$.
The normalized meson decay constant $F_\pi$ can then be obtained from the bare combination \cite{Shindler:2007vp}:
\begin{equation}
F_\pi = \frac{2 \mu_0  \braket{0 | P | \pi}_{\text{bare}} }{M_\pi^2}.
\end{equation}

\begin{table}[tp]
\centering
\begin{tabular}{c|c|c|c|c|c}
%\hline
  Ensemble& $L^3 \times T$ &$\beta$ & $am^s$ & $aM^s_\pi$ & $t_0^{\text{imp}}/a^2$ \\ \hline \hline
3A10 & $20^3 \times 36$ &\multirow{ 4}{*}{1.778}&-0.4040 & 0.2204(21) & 3.263(50)   \\ \cline{1-2} \cline{4-6} 
3A20 &$24^3 \times 48$ &&-0.4060 &  0.1845(14) & 3.491(32)  \\ \cline{1-2} \cline{4-6} 
3A30 &$24^3 \times 48$ &&-0.4070 & 0.1613(16)  & 3.740(39)  \\ \cline{1-2} \cline{4-6} 
3A40 &$32^3 \times 60$ &&-0.4080 &  0.1429(12)  & 3.855(27)  \\ \hline   \hline 
4A10 &$20^3 \times 36$ &\multirow{ 4}{*}{3.570} &-0.3725& 0.2035(14)& 3.494(45)  \\ \cline{1-2} \cline{4-6} 
4A20 &$24^3 \times 48$& &-0.3752&0.1805(7)& 3.565(26)  \\ \cline{1-2} \cline{4-6} 
4A30 &$24^3 \times 48$& &-0.3760& 0.1714(8)& 3.593(29) \\\cline{1-2} \cline{4-6} 
4A40 &$32^3 \times 60$& &-0.3780& 0.1397(8)& 3.723(23)  \\ \hline  \hline 
5A10 &$20^3 \times 36$& \multirow{ 4}{*}{5.969} &-0.3458& 0.2128(9)& 3.532(17)  \\ \cline{1-2} \cline{4-6} 
5A20 &$24^3 \times 48$& &-0.3490& 0.1802(6) & 3.614(18)  \\ \cline{1-2} \cline{4-6} 
5A30 &$24^3 \times 48$& &-0.3500& 0.1712(6)& 3.664(24) \\ \cline{1-2} \cline{4-6} 
5A40 &$32^3 \times 60$& &-0.3530&  0.1331(7)& 3.776(19)  \\ \hline  \hline 
6A10 &$20^3 \times 36$ &\multirow{ 4}{*}{8.974} &-0.3260& 0.2150(7) & 3.619(17) \\ \cline{1-2} \cline{4-6} 
6A20 &$24^3 \times 48$& &-0.3300&0.1801(5)& 3.696(17)   \\\cline{1-2} \cline{4-6} 
6A30 &$24^3 \times 48$& &-0.3311& 0.1689(7)& 3.721(15)  \\ \cline{1-2} \cline{4-6} 
6A40 &$32^3 \times 60$& &-0.3340& 0.1351(6)& 3.820(17) \\ \hline 
\end{tabular}
\caption{  Summary of our ensembles: $\beta$,  sea quark bare mass parameter, $m^s$, and sea pion mass $M^s_\pi$ .  We keep $c_{sw}=1.69$ throughout. }
\label{tab:ensembles}
\end{table}
\begin{table}[tp]
\centering
\begin{tabular}{c|c|c|c|c|c}

Ensemble & $ am_{\rm cr}$ & $a\mu_0$ & $aM^{\rm v}_\pi$ & $|am^{\rm v}_{pcac}|$ & $aF_\pi$ \\ \hline \hline
  3A10 &  -0.4214 &0.01107 & {0.2216(20)}&{0.0000(3)  }& 0.04405(41)  \\ \hline 
3A20 &  -0.4196&  0.00781& 0.1834(6) &0.0001(2)& 0.04023(24) \\ \hline 
3A30 &  -0.4187 &0.00632& 0.1613(11)&0.0008(2)&0.03678(33) \\ \hline 
3A40 &  -0.4163 &0.00513&0.1423(7)&0.0006(3)&0.03554(15) \\ \hline   \hline 
4A10 & -0.3875& 0.01030 & 0.2037(11)  & 0.0001(2)& 0.05131(37)\\ \hline 
4A20& -0.3865 & 0.00844&0.1803(9) & 0.0000(4) & 0.05037(26) \\ \hline 
4A30  & -0.3865 &0.00778 &0.1717(9)& 0.0001(4)  & 0.04913(31)\\ \hline 
4A40 & -0.3851& 0.00546  &0.1416(5)&0.0001(2)& 0.04608(15)  \\ \hline  \hline 
5A10& -0.3611& 0.01225  &0.2114(13) & 0.0003(4) &  0.06125(32) \\ \hline 
5A20&-0.3611& 0.00906 & 0.1799(10) & 0.0001(4)  & 0.05767(30) \\ \hline 
5A30&-0.3607& 0.00824 & 0.1706(13) & 0.0000(4) & 0.05647(40) \\ \hline 
5A40 &-0.3596& 0.00509  &0.1328(5)&0.0002(2)& 0.05278(18) \\ \hline  \hline 
6A10& -0.3415& 0.01298&0.2142(6) & 0.0003(2) & 0.06813(21)\\ \hline 
6A20 & -0.3414& 0.00956 & 0.1801(4)&0.0002(2)& 0.06435(25)\\ \hline 
6A30 & -0.3414& 0.00803  &0.1668(5)&0.0002(2)& 0.06278(24)\\ \hline 
6A40 &  -0.3409& 0.00542 &0.1342(4)&0.0000(1)& 0.05929(14) \\ \hline 
\end{tabular}
\caption{ Results obtained in the mixed action setup, with Wilson fermions on the sea and twisted mass in the valence sector. We use $c_{sw}=1.69$, as in the sea sector. }
\label{tab:results}
\end{table}

The results for the meson masses and decay constant in the mixed-action setup can be seen in Table \ref{tab:results}. We have achieved a  good tuning of $m_{\rm pcac}$ and the pseudoscalar masses are compatible within one or two sigma with their pure Wilson value (see Table \ref{tab:ensembles}). When the tuning to maximal twist is not perfect, we correct the bare quark mass (and thus $F_\pi$) as follows (see also \cite{Shindler:2007vp}):
\begin{eqnarray}
a\mu_0 & \rightarrow& a\mu_0 \sqrt{1 + \left( \frac{Z_A am_{\rm pcac}}{a\mu_0} \right)^2} , \\  
a F_\pi  &\rightarrow & a F_\pi \sqrt{1 + \left( \frac{Z_A am_{\rm pcac}}{a\mu_0} \right)^2} . 
\end{eqnarray}
where the axial normalization constant, $Z_A$, can be obtained non-perturbatively by matching the valence bare twisted mass with the PCAC mass measured in the sea sector:
\begin{equation}
\mu_0 = Z_A m^{ s}_{\rm pcac}, \text{ for } M_\pi^{\rm v} = M_\pi^s.
\end{equation}

\begin{figure*}[tp]
   \centering
   \subfigure[\ Ensembles with $N_c=3$ \label{fig:t0Nc3}]%
             {\includegraphics[width=0.45\textwidth,clip]{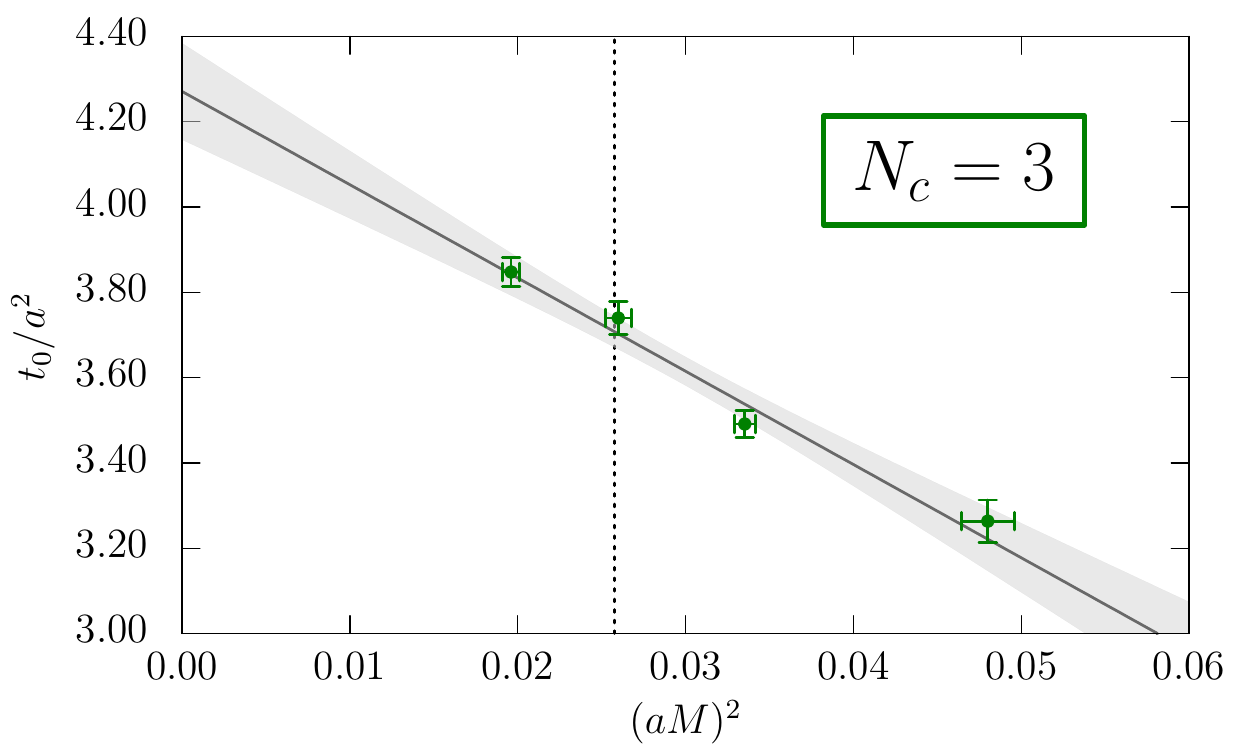}}\hfill
   \subfigure[\ Ensembles with $N_c=4$  \label{fig:t0Nc4}]%
             {\includegraphics[width=0.45\textwidth,clip]{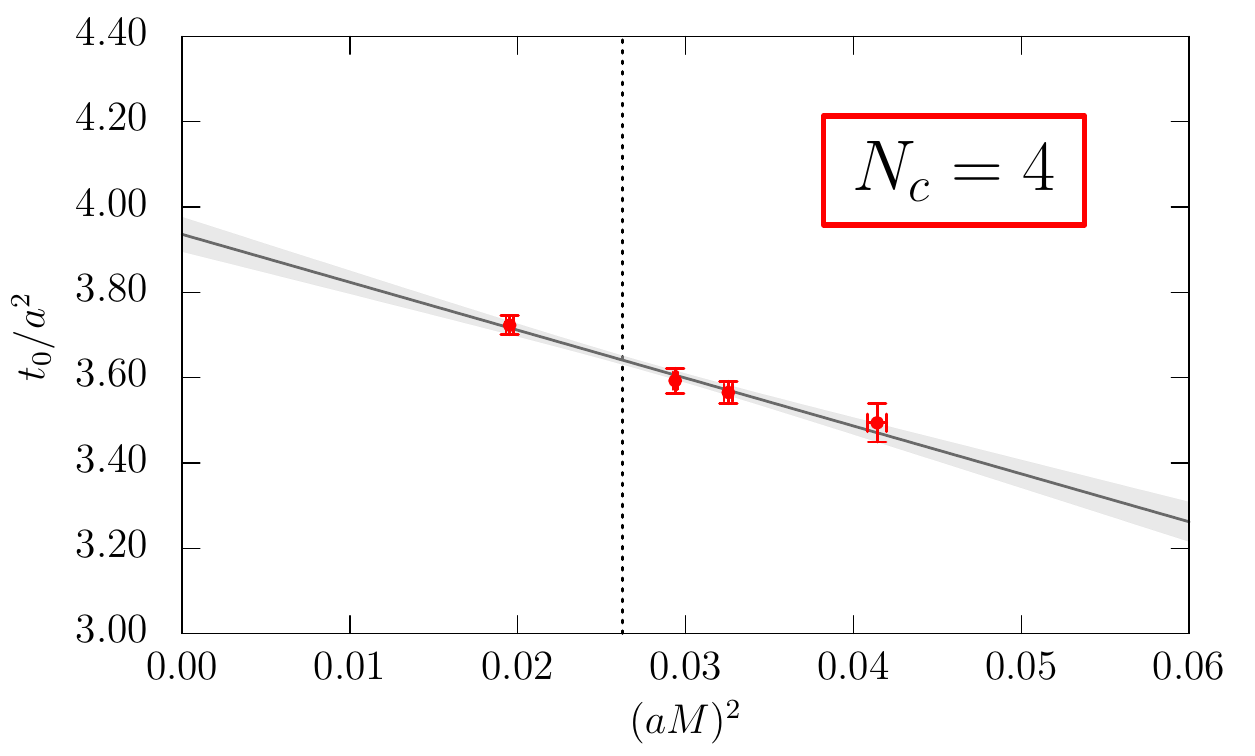}}
   \subfigure[\ Ensembles with $N_c=5$ \label{fig:t0Nc5}]%
             {\includegraphics[width=0.45\textwidth,clip]{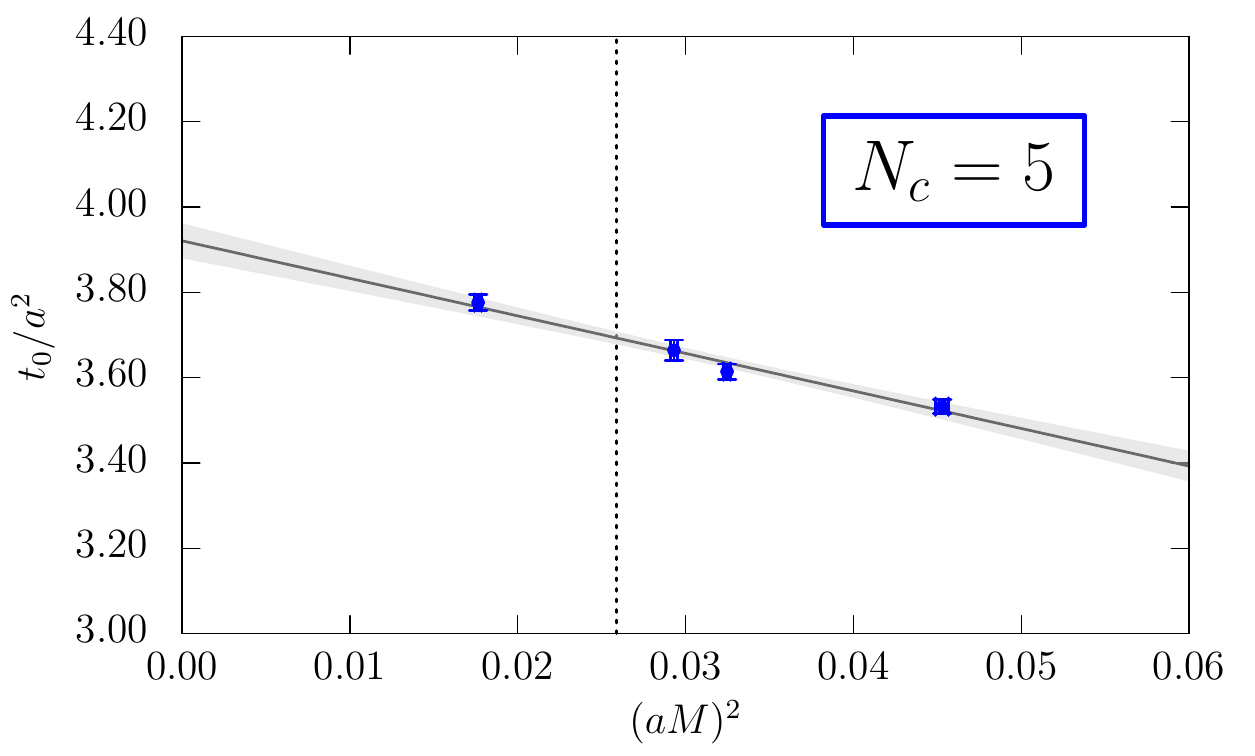}}\hfill
   \subfigure[\ Ensembles with $N_c=6$  \label{fig:t0Nc6}]%
             {\includegraphics[width=0.45\textwidth,clip]{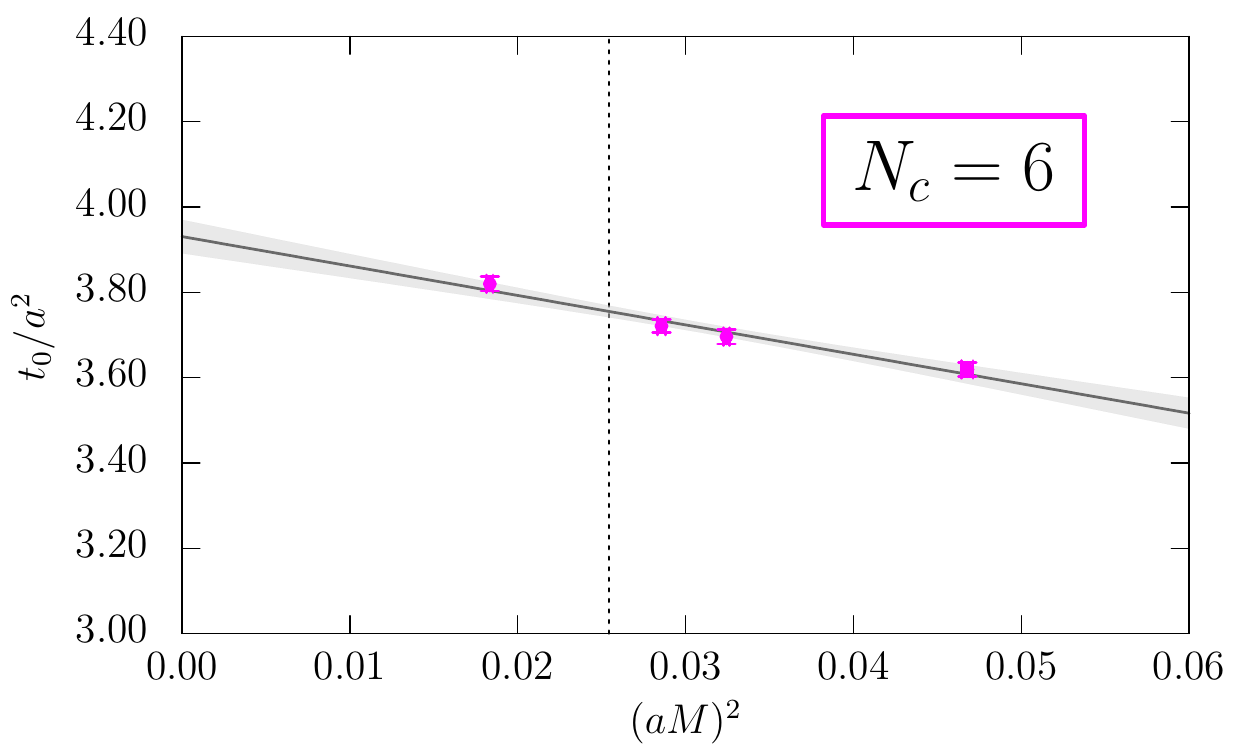}}
   \caption{Mass dependence of $t^{\rm imp}_0/a^2$. The vertical line corresponds to the value $M^2 = M^2_{\rm ref}$. }
   \label{fig:t0fig}
\end{figure*}
\section{Scale setting at Large $N_c$ \label{sec:scale}}
%In  Refs. \cite{Donini:2016lwz,Donini:2017rzi}, we performed a $N_c$ scaling ostudy in the quenched approximation,  the lattice spacing was kept constant by imposing the string tension to be constant for arbitrary values of $N_c$. 
The scale setting for different values of $N_c$ is performed using the gradient flow scale $\sqrt{8t_0}$, via the determination of $t_0/a^2$.
%, and our ensembles have been  the lattice spacing is constant across the different values of $N_c$. 
In QCD, with $N_c=3$, the standard definition of $t_0$ is:
\begin{equation}
\braket{ t^2 E(t)}\Big|_{t=t_0} = c =0.3.
\end{equation}
The leading dependence in $N_c$ is known \cite{Luscher:2010iy} in perturbation theory:
\begin{equation}
\braket{ t^2 E(t)} = \frac{3}{128 \pi^2} \frac{N_c^2-1}{N_c} \lambda_{\text{GF}}\left(q\right),
\end{equation}
where $\lambda_{\text{GF}}(q)$  is the gradient flow 't Hooft coupling at the scale $q=1/{\sqrt{8t}}$. 
%The latter is related to the $\overline{\text{MS}}$ coupling at 
%NLO in perturbation theory as  \cite{Luscher:2010iy,Harlander:2016vzb}:
%\begin{equation}
%\lambda_{\text{GF}}\left(q\right) = \lambda_{\overline{\text{MS}}}\left(q\right)  \left( 1+  \frac{c_1}{4\pi} \lambda_{\overline{\text{MS}}}(q) + O(\lambda_{\overline{\text{MS}}}^2) \right),
%\label{eq:t0N}
%\end{equation}
%where $c_1=0.36593+0.0075 {N_f}/{N_c}$. 
% From Eq. \eqref{eq:t0N}, we see that the $N_f$ dependence arises at 1 loop and with a very small coefficient.
Hence, as in Ref.  \cite{Ce:2016awn}, we will generalize $t_0$ to an arbitrary $N_c$ as:
\begin{equation}
\braket{ t^2 E(t)}\Big|_{t=t_0} = c(N_c)={3\over 8} \frac{N_c^2-1}{N_c} c(3). \label{eq:t0N2}
\end{equation}
Notice that the choice here is not unique. In particular, one could choose another coupling in a different scheme (such as $\overline{\text{MS}}$ ), and this would induce corrections at order O($N_f/N_c$) in dimensionful quantities. 
%From Eq. \eqref{eq:t0N}, we see that the $N_f$ dependence arises at the one-loop level and with a very small coefficient, and thus the subleading $N_c$ corrections induced by our particular choice of scale setting are expected to be small.
%which at NLO is true up to a very small $N_f/N_c$ corrections ($\sim 0.5 \%$).

We also need the value of $t_0$ in physical units. This is known from lattice simulations for $N_f=2$  \cite{Bruno:2013gha, Sommer:2014mea} and $N_f=3$ \cite{Bruno:2016plf} degenerate quarks and at a reference pion mass $M_{\rm ref}=420\text{ MeV}$:
\begin{align}
\begin{split}
\sqrt{t_0}\Big|^{N_f=2}_{M_{\rm ref}} = 0.1470(14) \text{ fm,} \\ \sqrt{t_0}\Big|^{N_f=3}_{ M_{\rm ref}} = 0.1460(19) \text{ fm} 
\end{split}
\end{align}
We can use these to perform a linear extrapolation to $N_f=4$, motivated by the weak $N_f$ dependence:
\begin{align}
\sqrt{t_0}\Big|^{N_f=4}_{ M_{\rm ref}} = 0.1450(39) \text{ fm}.
\end{align}
Our scale setting condition involves therefore the dimensionless quantity
\begin{equation}
(M_\pi \sqrt{t_0}) \Big|_{M_{\rm ref}} = 0.3091(83). \label{eq:t0Mref}
\end{equation}
%Note that the Eq.~(\ref{eq:t0N}) is consistent with having a slightly smaller value for $t_0$ as $N_f$ increases.

In order to reduce discretization errors we have performed a tree level improvement of $t_0$.  In Ref. \cite{Fodor:2014cpa}, lattice perturbation theory is used to improve $\braket{t^2 E(t)}$ and thus, $t_0$. The prescription is: 
\begin{align}
\braket{t^2 E(t)}_a =  \braket{t^2 E(t)}_{\text{imp}}  \Bigg[ &1+ \sum_{n} C_{2n} \left(\frac{a^{2}}{t}\right)^n   \Bigg], \label{eq:Et2imp}
\end{align}
where the coefficients $C_{2n}$ depend on the gauge action, the flow action and the definition of $E(t)$ (clover or plaquette). 
%We use Iwasaki gauge action for the simulation and plaquette action for the flow. For the observable, we will use the clover definition for the central value and the difference with from the two definitions ?? will be considered a remaining discretization effect.  For the clover definition, the coefficients are:
The coefficients for the Iwasaki gauge action, the plaquette action for the flow and the clover definition of $E(t)$ are:
\begin{align}
\begin{split}
C_2 = -0.262333, \ C_4 = 0.0936935, \\  C_6 = -0.048002, \ C_8=0.0320211.
\end{split}
\end{align}
The numerical results after the improvement, $t_0^{\rm imp}/a^2$, are shown in Table \ref{tab:ensembles}. 
\begin{table}[H]
\centering
\begin{tabular}{c|c|c}
$N_c$ & $\left.t_0/a^2\right|_{M_{\text{ref}}}$  &$a$ ($\times 10^{-2}$ fm) \\ \hline\hline
3     & $3.71(4)(7)_{\small t_0}(12)_a(3)_L$ &$7.53(4)(19)_{t_0}(12)_{a}(3)_L$    \\ \hline
4     & $3.64(1)(3)_{\small  t_0}(12)_a(3)_L$ &$7.60(1)(20)_{t_0}(12)_{a}(3)_L$  \\ \hline 
5     & $3.69(2)(3)_{\small t_0}(12)_a(3)_L$ &$7.54(2)(20)_{t_0}(12)_{a}(3)_L$ \\ \hline
6     & $3.76(1)(2)_{\small t_0}(12)_a(3)_L$ &  $7.48(1)(20)_{t_0}(12)_{a}(3)_L$  \\ \hline\hline
\end{tabular}
\caption{Results for the $\left.t_0/a^2\right|_{M_{\text{ref}}}$ and the lattice spacing as a function of $N_c$. The first error is statistical, the second comes from the uncertainty in $t_0$ in physical units, the third stems from the difference in the definitions of $E(t)$ after improvement, and the fourth are finite volume effects estimated from Ref. \cite{Fodor:2012td}.}
\label{tab:scalesetting}
\end{table}

Finally, Eq.~\eqref{eq:t0Mref} requires $t_0$ at  $M_{\rm ref}$. The mass dependence of $t_0$ has been studied in chiral perturbation theory in Ref. \cite{Bar:2013ora}. For degenerate flavours it is given by
\begin{equation}
t_0  =   t_0^{\chi} \left(1 + k \ M^2 \right)  + {\mathcal O}(M^4),
\end{equation} 
where $k \propto 1/(F_\pi)^2 = O(1/N_c)$ and so the chiral dependence is suppressed in $N_c$. We have performed  accordingly a linear fit in $M^2$ to extract the reference value. The mass dependence of $t^{\text{imp}}_0$ for the different values of $N_c$ can be seen in Fig. \ref{fig:t0fig}. As expected, the slope is suppressed with $N_c$. The results of the scale setting can be seen in Table \ref{tab:scalesetting}, where we also include the systematic uncertainties. The leading uncertainty comes from the error on the value of $t_0$ in physical units,  the discretization error is estimated from the difference in two definitions of $E(t)$ after improvement, and the  finite volume systematic error is estimated from Ref. \cite{Fodor:2012td}. As it can be seen, the scale setting yields a uniform lattice spacing for all the values of $N_c$. From now on, we will quote our results in terms of the lattice spacing $a = 0.0754 \text{ fm}$, corresponding to $N_c=5$.

%  %For the physical quantities, we will use dimensionless products in terms of $\left.t_0/a^2\right|_{M_{\text{ref}}}$ henceforth.

\section{Chiral Perturbation Theory Fits}
\label{sec:fits}

The results for $M_\pi$ and $F_\pi$ in the mixed-action setup are presented in Table \ref{tab:results}.  We want to compare these results to the expectations in ChPT described in Sec. \ref{sec:chpt} in order to the extract the LECs and study their $N_c$ scaling. 

Before addressing the fits, we need to explain some technical issues regarding the finite volume effects, the renormalization scale and the fitting strategy. We then perform fits at a fixed value of $N_c$ to test the {ans\"atze} for the $N_c$ scaling of the LECs in Eqs. \ref{eq:lonc} and \ref{eq:lecsnc}. After that, we perform simultaneous chiral and $N_c$ fits. We present a selection of relevant results for the latter, and conclude the section with a discussion on systematic errors.

 %We wil now describe the fit strategy and our results.
\subsection{Finite volume effects}

Our ensembles have $M_\pi L > 3.8$ in all cases so we expect finite volume effects to be small and suppressed as $1/N_c$. Still, we find that for the decay constant they can be of $O(1\%)$ and thus we correct them as \cite{Gasser:1986vb, Colangelo:2005gd}:
\begin{align}
M_\pi(L) &= M_\pi \bigg[ 1+ \frac{1}{2 N_f}  \xi \ \bar{g}_1(M_\pi L) + O(\xi^2)\bigg], \\
F_\pi(L) &= F_\pi \bigg[ 1- \frac{N_f}{2 }  \xi \ \bar{g}_1(M_\pi L) + O(\xi^2)\bigg], \label{eq:FV}
\end{align}
with  $\xi \equiv\frac{M_\pi^2}{(4 \pi F_\pi)^2}$, while $\bar{g}_1(x)$ is given by
\begin{equation}
\bar{g}_1(x) \xrightarrow{x \gg 1}  \frac{24}{x} K_1(x) \sim \frac{24 \sqrt{2}}{\sqrt{\pi}} \frac{e^{-x}}{x^{3/2}}.
\end{equation}
We will use the corrected results for the analysis.
\subsection{Renormalization scale}
The NLO couplings are usually defined at $\mu=4 \pi F$ or at the $\rho$ mass, $\mu=M_\rho$. Still, in the context of the large $N_c$ expansion these are two very different choices, since the former scales with $\sqrt{N_c}$, deviating from the physical cutoff
of the chiral effective theory, which is expected to be set by the lighter resonances, such as the $\rho$. The scale $\mu= 4 \pi F$ is instead the scale at which ChPT breaks down, which for large enough $N_c$ is much higher than the scale at which new resonances appear. In the context of large $N_c$, it is therefore sensible to choose a renormalization scale more closely related
to the physical cutoff that does not scale with $N_c$.  Keeping the scale related to $4\pi F$, however, has some advantages for fitting, so we choose:
\begin{eqnarray}
\mu^2 = {3 \over N_c} (4 \pi F)^2,
\end{eqnarray} 
which has no leading dependence on $N_c$. Using this scale, the NLO predictions can be conveniently written as:
\begin{eqnarray}
F_\pi &=& F \Bigg[1   - 2 \xi \log\left({N_c \over 3} \xi\right)    + 64 \pi^2 \xi L_F(\mu)  \Bigg], \label{eq:fnloxi}\\
\frac{M_\pi^2}{m} &=& 2 B   \Bigg[1   + \frac{1}{4} \xi \log\left({N_c \over 3} \xi\right)    + 128 \pi^2  \xi L_M(\mu)  \Bigg],
\label{eq:mnloxi}
\end{eqnarray}
where $m=\mu_0$, the bare twisted mass. Note that in this expression $B$ is bare, since the quark mass is also bare. The value of the non-singlet pseudoscalar normalization constant, $Z_P$, is thus needed.
%Both need a multiplicative renormalization factor that cancels in the product $\mu_0 B$. 

\subsection{Fitting strategy}

Some care is needed to perform the fits in Eqs. (\ref{eq:fnloxi}) and (\ref{eq:mnloxi}). The complication comes from the fact that both coordinates, $(x,y)= (\xi, F_\pi)$ or $(x,y)= (\xi, M^2_\pi/\mu_0)$ have correlated errors. In particular the Ordinary Least Square (OLS) method is not appropriate, since it assumes no errors in $x$ coordinate. An alternative approach is the York Regression (YR) \cite{YORK}, in which the $\chi^2$ function is:
\begin{equation}
\chi^2 = \sum_i  \min_{\delta_i} \left[ {\mathbf R}^T_i  V^{-1}  {\mathbf R}_i \right],
\end{equation}
where we have defined the two-dimensional vectors:
\begin{equation}
{\mathbf R}_i(\delta_i) \equiv \left( f( x_i + \delta_i ) - y_i, \delta_i  \right),
\end{equation}
where $f$ is the fitting function, and $V$ is the $x,y$-covariance matrix, estimated using bootstrap samples. { In order to account for autocorrelations, we vary the block-size of the bootstrap samples. We find that blocks of $\sim 20$ units of Montecarlo are sufficient, and we do not observe a clear $N_c$ dependence. We also estimate all the errors of the fit parameters via bootstrap resampling.}

\subsection{Fit results at fixed $N_c$}

\begin{figure}[h!]
\centering % \begin{center}/\end{center} takes some additional vertical space
\includegraphics[width=.475\textwidth]{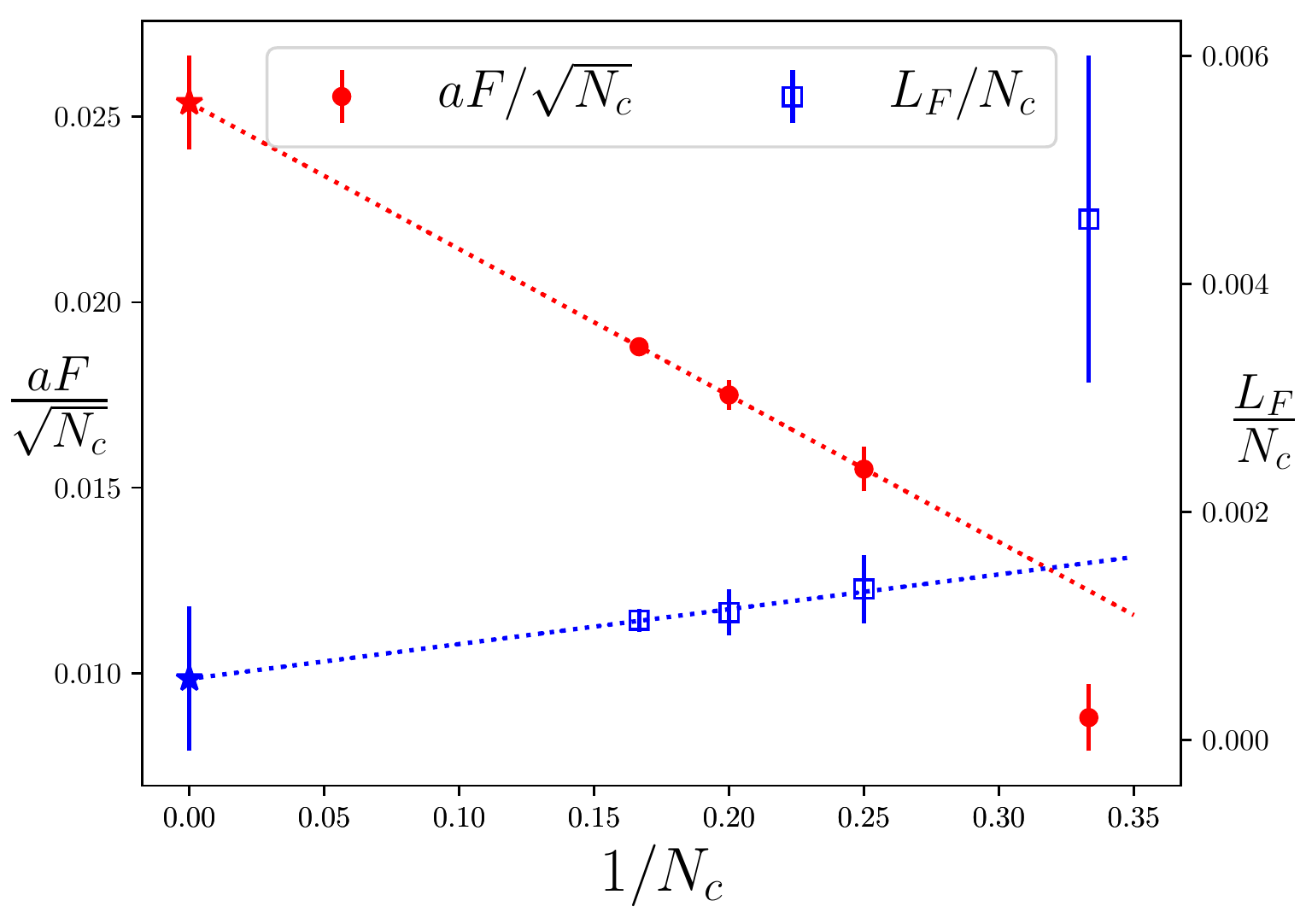}
\caption{\label{fig:LECsFNLO} $N_c$ dependence of $F/\sqrt{N_c}$ (red) and $L_F$ (blue). The dotted lines are the best fits to Eqs.~(\ref{eq:lonc}) and (\ref{eq:lecsnc}) excluding the data points at $N_c=3$.}
\end{figure}

\begin{figure}[h!]
\centering % \begin{center}/\end{center} takes some additional vertical space
\includegraphics[width=.475\textwidth]{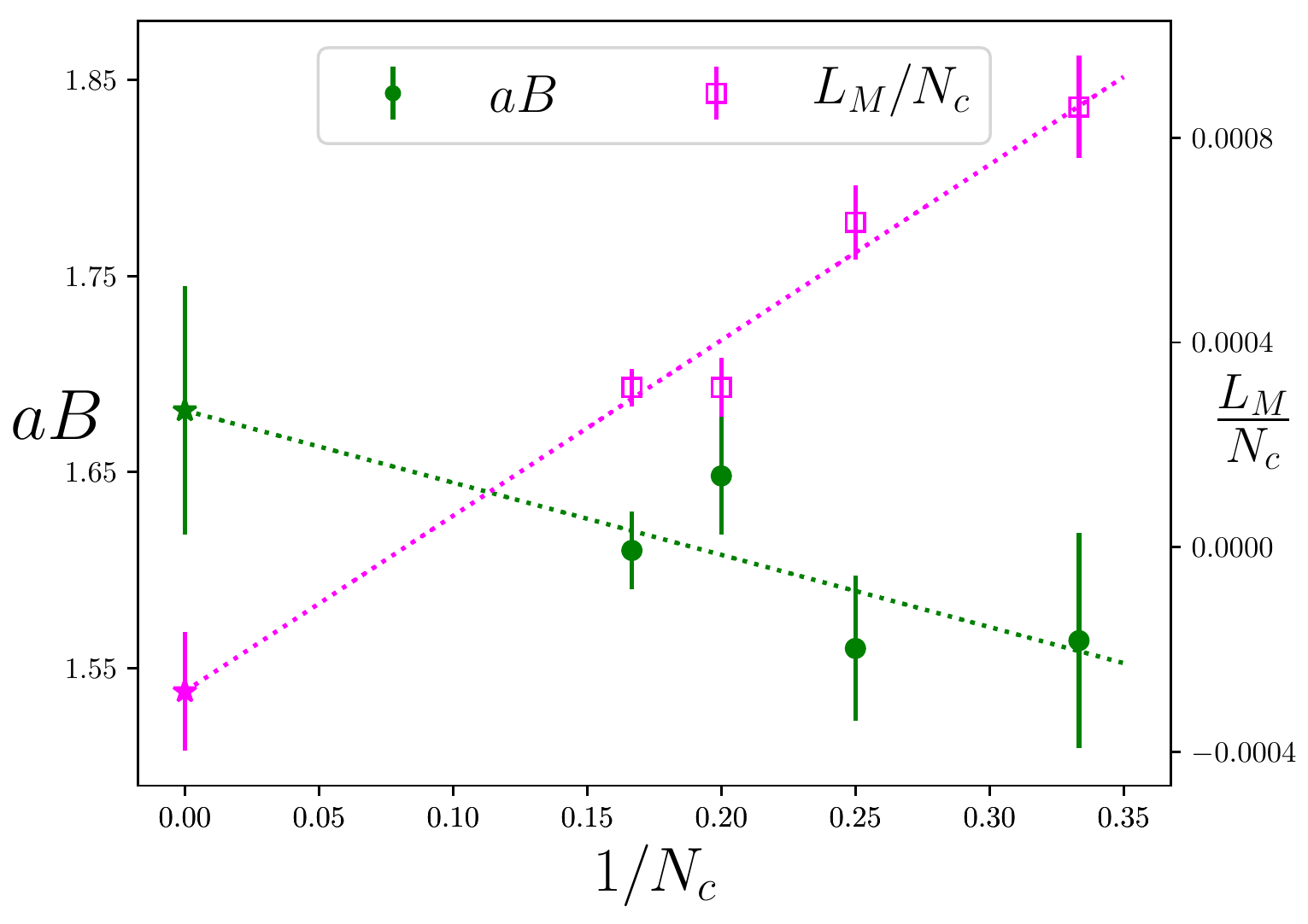}
\caption{\label{fig:LECsMNLO} $N_c$ dependence of $B$ and $L_M$. The dotted lines are the best fits to Eqs.~(\ref{eq:lonc}) and (\ref{eq:lecsnc}) including all points. }
\end{figure}

First we consider each $N_c$ separately and perform a fit of the data points to extract $F, L_F(\mu)$ and  $B, L_M(\mu)$.  The NLO fit results for these quantities are shown respectively in Tables~\ref{tab:separateF} and Tables~\ref{tab:separateM}.  The 
$N_c$ dependence of the LECs is shown in Figs.~\ref{fig:LECsFNLO} and ~\ref{fig:LECsMNLO}. It can be seen that the scaling is well described by leading and subleading $N_c$ corrections for $N_c=4-6$, while there seems to be significant $1/N_c^2$ corrections for $N_c=3$ in the case of  $F$ and $L_F$. In the case of $B$ and $L_M$ errors are larger and there is no sign of $1/N_c^2$. Interestingly, the data suggest that the large $N_c$ limit of $L_M \sim 0$.

\begin{table}[h!]
\centering
\begin{tabular}{c|c|c|c}
$N_c$ & $aF/\sqrt{N_c}$ & $L_F/N_c$      & $\chi^2/dof$ \\ \hline \hline
3     & 0.0088(9)   & 0.0046(14) & 0.7/2       \\ \cline{1-4}
% & 0.0087(13)   & 0.014(9) & 0.62/1       \\ \hline
4     & 0.0155(6)   & 0.0013(3) & 3.9/2       \\ \hline
5     & 0.0175(4)   & 0.0011(2) & 2.2/2       \\ \hline
6     & 0.0188(2)   & 0.0011(1)  & 0.4/2       \\ \hline
%$\infty$ \cite{Bali:2013kia}    & 0.0213(21)   & 0.00063(9) $ \cdot N_c$  & ---       \\ \hline
\end{tabular}
\caption{NLO Fits for $F_\pi$ for separate values of $N_c$.  }
\label{tab:separateF}
\end{table}
%For the mass, we try differnt fits
\begin{table}[h!]
\centering
\begin{tabular}{c|c|c|c}
 $N_c$ & $aB$ & $L_M/N_c$      & $\chi^2/dof$ \\ \hline  \hline
 3    & 1.564(55)   & 0.00086(10) & 10.2/2       \\ \cline{1-4}
%      & 1.348(81)   & 0.00404(64) & 1.28/1       \\ \hline
  4     & 1.560(37)   & 0.00064(7) & 1.4/2       \\ \hline 
  5     & 1.648(30)   & 0.00031(6) & 0.1/2       \\ \hline 
 6     & 1.610(20)   & 0.00031(4)  & 9.5/2       \\ \hline
%$\infty$ [Bali]    & 0.0213(21)   & 0.00063(9) $ \cdot N_c$  & ---       \\ \hline
\end{tabular}
\caption{Fits for $M_\pi$ for separate values of $N_c$}
\label{tab:separateM}
\end{table}

%We  therefore truncate the large $N_c$ behaviour of the LECs to NLO, as in eqs.~\eqref{eq:lonc} and \eqref{eq:lecsnc}, in a global fit that includes all $N_c$ for the mass dependence, while we include also $1/N_c^2$ corrections for that of the decay constant. 

\subsection{Simultaneous chiral and $N_c$ fits}

We now consider a global fit including several data points at different values of $N_c$. We first perform a $SU(N_f)$-NLO fit to the subset $N_c=4-6$, including leading and subleading $N_c$ corrections for all the LO and NLO LECs, as parametrized in Eqs.~\eqref{eq:lonc}  and \eqref{eq:lecsnc}. We linearize the fit by considering the following parametrization
\begin{align}
\begin{split}
F_\pi &= \sqrt{N_c} \left(F_0+ {F_1\over N_c}\right) \Bigg[1   - 2 \xi \log\left({N_c \over 3} \xi\right)  \Bigg]   \\
&+ 64 \pi^2 \xi \sqrt{N_c} \left(N_c (F L_F)^{(0)} + (F L_F)^{(1)} \right) , \label{eq:fnlo}
\end{split}
\\ \begin{split}
\frac{M_\pi^2}{m} &= 2 \left(B_0 + {B_1 \over N_c}\right)   \Bigg[1   + \frac{1}{4} \xi \log\left({N_c \over 3} \xi\right)  \Bigg]  \\
&+ 256 \pi^2  \xi \left(N_c (B L_M)^{(0)} + (B L_M)^{(1)}\right), 
\end{split}
\label{eq:mnlo}
\end{align}
where $(F L_F)^{(0)} \equiv F_0 L_F^{(0)}$, while $(F L_F)^{(1)} \equiv F_1 L_F^{(0)}+ L_F^{(1)} F_0$, and $(B L_M)^{(0)} \equiv B_0 L_M^{(0)}$ and $(B L_M)^{(1)} \equiv B_1 L_M^{(0)}+ B_0 L_M^{(1)}$.

Secondly, we consider the $U(N_f)$-NNLO  expansion, since we have checked that the $U(N_f)$-NLO expressions fit the data very poorly. We also linearize the fit by considering 
the following fitting functions:
\begin{eqnarray}
F_\pi &=&  \sqrt{N_c} \left(F_0+ {F_1\over N_c} + {F_2 \over N_c^2}\right)  \Bigg[1   - 2 \xi \log \left({N_c \over 3} \xi\right) \Bigg]\nonumber\\
&+& 64 \pi^2 \xi \sqrt{N_c} \left(N_c (FL_F)^{(0)} + (FL_F)^{(1)}\right)   \\  
%   &+&  N_c^2 \left(16 \pi^2 \xi \right)^2  \left(24  (L^{(0)}_F)^2  -  32 L^{(0)}_M L^{(0)}_F + C^{(0)}_F\right)   \nonumber\\
   &+&  N_c^2 \sqrt{N_c} \left(16 \pi^2 \xi \right)^2  K^{(0)}_F,  \nonumber \label{eq:FpiNNLOleut}
\end{eqnarray}
\begin{eqnarray}
\frac{M_\pi^2}{m} &=& 2 \left( B_0 + {B_1\over N_c} + {B_2 \over N_c^2}\right) \Bigg[  1   +  \frac{1}{4} \xi \log \left({N_c \over 3} \xi\right)  \nonumber\\
&+&  -\frac{1}{4}  \left(\xi+{a_0\over N_c^2}\right) \log \left({N_c \over 3} \left(\xi+{a_0\over N_c^2}\right)\right) \Bigg]  \\
&+&256 \pi^2 \xi \left(N_c  (B L_M)^{0} + (B L_M)^{(1)}\right) \nonumber\\
  &-&  64  N_c^2 \left(16 \pi^2  \xi\right)^2 K_M^{(0)}, \nonumber \label{eq:MpiNNLOleut}
\end{eqnarray}
 where
\begin{eqnarray}
a_0 \equiv N_c^2 {M_0^2\over (4 \pi F)^2},
\end{eqnarray}
and $M_0^2$ is given by the Witten-Veneziano formula for the $\eta'$ mass valid in the large $N_c$ limit (see Eq. \ref{eq:WV}). 
We use the result for the topological susceptibility from Ref.~\cite{Ce:2016awn}, 
\begin{eqnarray}
t_0^2 \chi_t = 7.03 (13) \cdot 10^{-4}. 
\end{eqnarray}
We convert to lattice units using the value of $t_0/a^2$ in the previous section and substitute $F\rightarrow \sqrt{N_c} F_0$, as extracted from the global $F_\pi$ fit. 
We find $a_0 \sim 6.5$, a value we fix in the fit.

In summary we compare the following fits:
\begin{enumerate}[label=\roman*)]
\item Fit 1: $SU(N_f)$-NLO fit to Eq.~\eqref{eq:fnlo} and \eqref{eq:mnlo}  including the data subset $N_c=4$-6. %For the decay constant we will include constraints from the large $N_c$ quenched results from Ref. \cite{Bali:2013kia}.
\item Fit 2: $U(N_f)$-NNLO expansion fit  to Eqs.~\eqref{eq:FpiNNLOleut} and \eqref{eq:MpiNNLOleut} including the full data set.
%\item Fit 2: as fit 1 but including also the data set $N_c=3$ and adding a sub-subleading $N_c$ term in $F$ and $L_F$. 
%\item Fit 3: NNLO fit to eqs.~\eqref{eq:FpiNNLO} and  \eqref{eq:MpiNNLO}  with parametrizations of LECs \eqref{eq:lonc}-\eqref{eq:lecsnc} and $K_{F/M} = N_c K_{F/M}^{(0)}$, including the data subset $N_c=4$-6.
%\item Fit 4: as fit 3, but including also the data set $N_c=3$ and adding a sub-subleading $N_c$ term in $F$ and $L_F$.
\end{enumerate}
%We also include constraints on $F_0$ and $L_F^{(0)}$ from the quenched results from Ref. \cite{Bali:2013kia}, since the quenched approximation should correctly capture the leading behaviour in large $N_c$. These constraints are included as additional terms in the $\chi^2$:
%\begin{equation}
%\chi^2 \rightarrow \chi^2 +\left(\frac{F_0-F_0|_{\rm Bali} }{ \sigma_{F_0}}\right)^2 +\left(\frac{L_F^0 - L_F^0 |_{\rm Bali}}{\sigma_{L_F^0}}\right)^2.
%\end{equation}

\begin{figure*}[tpb]
   \centering
   \subfigure[\label{fig:Fpi} Chiral fits for the decay constant.]%
             {\includegraphics[width=0.475\textwidth,clip]{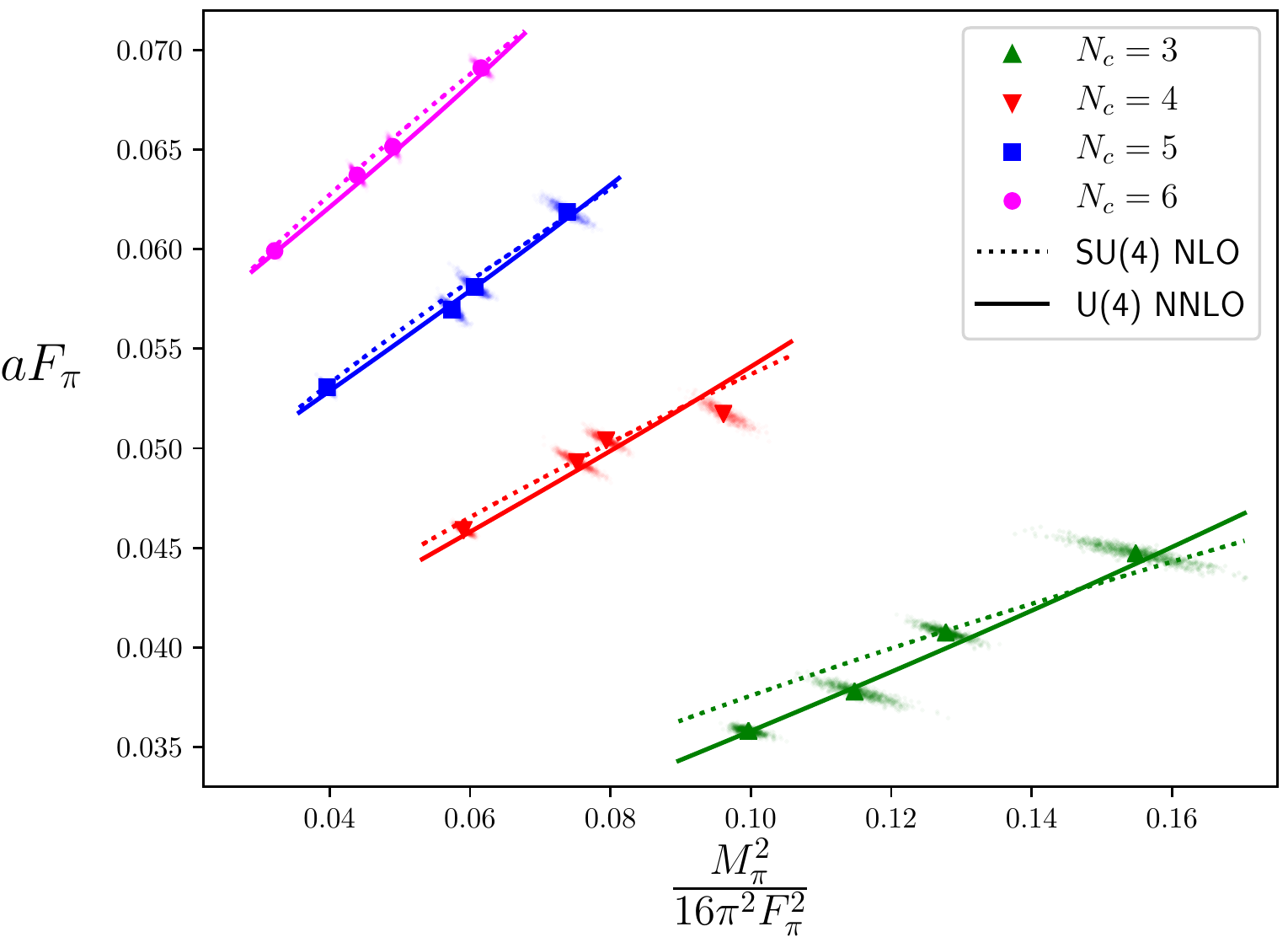}}\hfill
   \subfigure[ \label{fig:Mpi} Chiral fits for the meson mass.]%
             {\includegraphics[width=0.475\textwidth,clip]{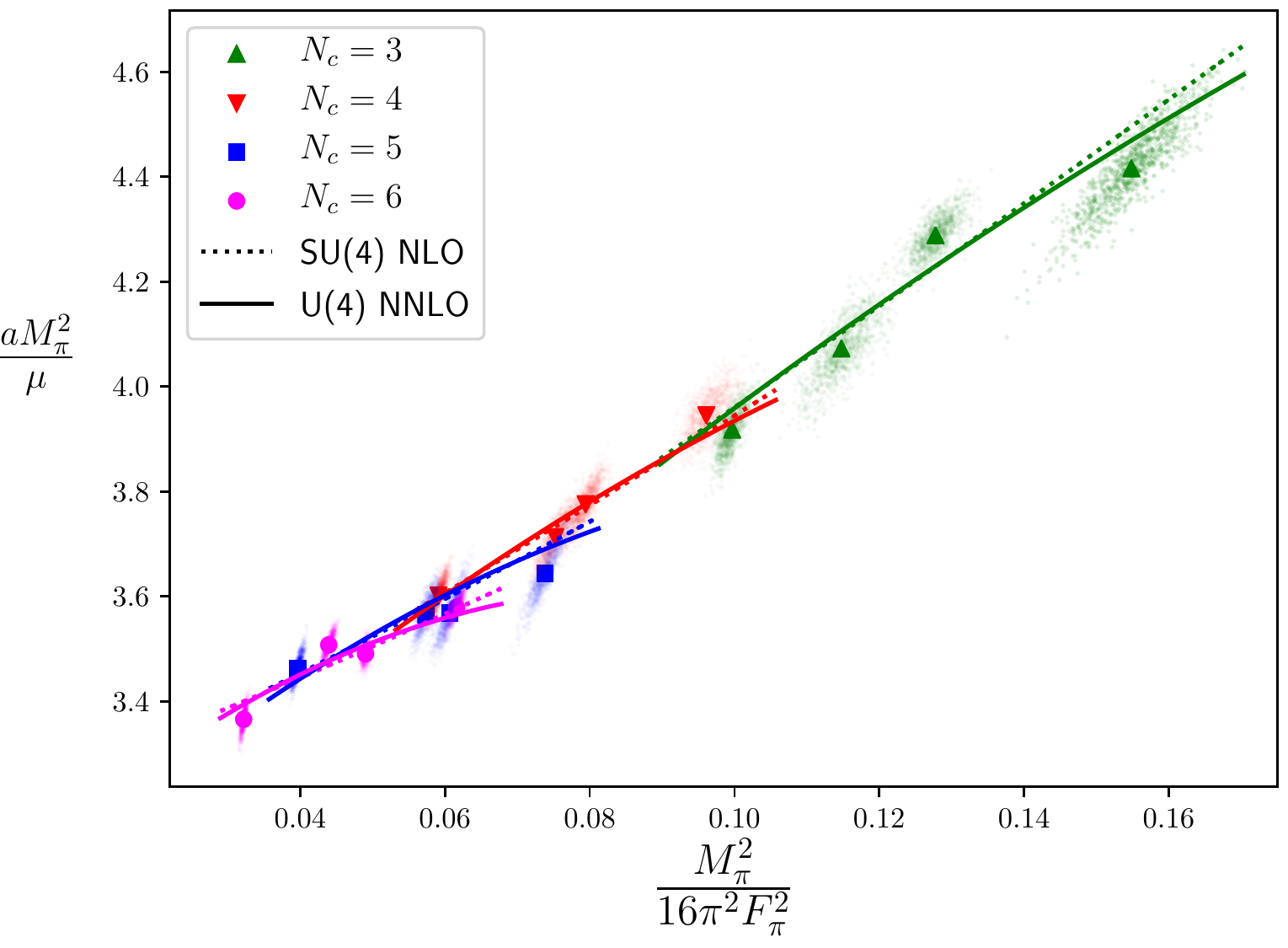}}
  
   \caption{Data and NLO/NNLO fits for the decay constant and meson mass. The central value is shown together with the bootstrap samples used for fitting. The results include finite-volume corrections as in Eq.~(\ref{eq:FV}) }
   \label{fig:chiralfits}
\end{figure*}

The results for the fitted parameters in the global fits  are shown in Tables \ref{tab:allF} and \ref{tab:allM}, and the quality of the fits is shown in Figs.~\ref{fig:Fpi} and \ref{fig:Mpi}. We also quote in Table \ref{tab:allLECs} the results for the NLO LECs from these fits. Errors are large, but there are significant correlations between the parameters as can be seen in Fig. \ref{fig:corr}.

\subsection{Selected results}

We will now quote some results that can be inferred from our fits. We first focus on the decay constant in the chiral limit. Using $a= 0.0754(23)$ fm, we get from our fits at fixed $N_f=4$:
\begin{align}
{\rm Fit 1}: {F\over  \sqrt{N_c}} &=  \left(67(3) - 26(4){N_f\over N_c} \right)(3\%)^a{\rm MeV},  \label{eq:fitFpiMev}  \\
{\rm Fit 2}: {F \over \sqrt{N_c}} &= \left(70(2)- 22(5){N_f\over N_c} -  {86(37)\over N_c^2}\right)(3\%)^a {\rm MeV},  \nonumber
\end{align}
where the $N_f$ dependence assumed is the expected one as discussed in sec.~\ref{sec:chpt}. Note that no $N_f$ dependence is assumed in the  $1/N_c^2$ terms. 
The first error is just the one obtained from the fits in Table \ref{tab:allF} and the second error of $3\%$ is the one corresponding to the lattice spacing determination. For two- and three-flavour QCD we get:
\begin{align}
\begin{split}
{\rm Fit 1}: & F^{N_c=3, N_f=2} = 86(3)  \text{ MeV}, \\
                  & F^{N_c=3, N_f=3} = 71(3)  \text{ MeV}, \label{eq:FNf1}
\end{split} \\
\begin{split}
{\rm Fit 2}: & F^{N_c=3, N_f=2} = 81(7)  \text{ MeV},  \\
                  & F^{N_c=3, N_f=3} = 68(7)  \text{ MeV},\label{eq:FNf2}
\end{split}
\end{align}
where we have taken into account the correlations between the different terms in Eq. (\ref{eq:fitFpiMev}), and we have assumed no $N_f$ dependence on the last term of the Fit 2. These results are in perfect agreement with phenomenological determinations: 
\begin{align}
\begin{split}
F^{N_f=2} &= 86.2(5) \text{ MeV} \text{ in Ref. \cite{Colangelo:2003hf}},   \\
F^{N_f=3} &\simeq 71.1  \text{ MeV} \text{ in Ref. \cite{Ananthanarayan:2017yhz}}, 
\end{split}
\end{align}
 and also lattice results (see Ref. \cite{Aoki:2019cca}). In addition, we can compare to previous results in the large $N_c$ limit in the quenched approximation:
\begin{equation}
{F\over  \sqrt{N_c}} {\bigg \rvert}_{N_c \to \infty} = 56(5) \text{ MeV} ,  \text{  Ref.  \cite{Bali:2013kia}}.
\end{equation}
This value is $2\sigma$ away from the results in Eq.~(\ref{eq:fitFpiMev}). This discrepancy may be explained however with the lack of non-perturbative normalization constant and discretization effects, which in their case are of $O(a)$.

Regarding the coupling, $B\equiv \Sigma/F^2$, we do not have a non-perturbative value of $Z_P$, up to this factor we get:
\begin{eqnarray}
{\rm Fit 1}: {\Sigma \over F^2} &=& Z_P \left(  1.70(11)- 0.12(12) {N_f\over N_c} \right),\label{eq:fitMpiMev} \\
{\rm Fit 2}: {\Sigma \over F^2} &=& Z_P \left(1.72(7) - 0.45(37){N_f\over N_c} -  {1.8(1.5) \over N_c^2}\right). \nonumber
\end{eqnarray}
%\begin{table}[h!]
%\centering
%\begin{tabular}{c||c|c|c|c}
% $N_c$ & 3 &4 & 5 & 6 \\ \hline
% $Z_P$    & 0.555 &0.584&0.602&0.614       \\ 
%\end{tabular}
%\caption{Values of $Z_P$ for different $N_c$, using the 1-loop perturbative results of Ref. \cite{Alexandrou:2012mt} for our particular set-up.}
%\label{tab:Zp}
%\end{table}
%Using the perturbative results for $Z_P$, shown in Table \ref{tab:Zp}, we get for $\hat B = B Z_P$ ({\bf isn't it the inverse}):
From Ref. \cite{Alexandrou:2012mt}, we can obtain the the 1-loop perturbative result for the normalization constant: 
\begin{equation}
Z_P(N_c=3) = 0.555,
\end{equation}
 which at the order we are working is independent of $N_f$. With this, we obtain for $N_c=3$:
\begin{align}
\text{Fit 1: } & \  N_f=4&\  \longrightarrow& \ \frac{\Sigma}{F^2}= 2.26(11) (7)^a \text{ GeV},  \\
& \  N_f=3&\  \longrightarrow& \ \frac{\Sigma}{F^2}= 2.31(5) (7)^a \text{ GeV}, \\
& \  N_f=2&\  \longrightarrow& \ \frac{\Sigma}{F^2}= 2.35(3) (7)^a \text{ GeV}, 
\end{align}
where the first error is systematic, the second comes from the scale setting, and we omit any systematic errors regarding the normalization constant. Combining these results with the ones in Eqs.~(\ref{eq:FNf1}) and (\ref{eq:FNf2}), we obtain:
\begin{align}
\Sigma^{1/3}(N_f=2) &= 257(2)(9)^a \text{ MeV}, \\
\Sigma^{1/3}(N_f=3) &= 223(4)(8)^a \text{ MeV},
\end{align}
which is compatible within $1\sigma$ with the numbers quoted in Ref.~\cite{Aoki:2019cca}.
We can also consider the ratio of condensates for $N_f=2$ and $N_f=3$, where the $Z_P$ factor drops (up to subleading $N_f$ dependence):
\begin{equation}
\frac{\Sigma(N_f=2)}{\Sigma(N_f=3)} = 1.49(10),
\end{equation}
which shows good agreement with the prediction  
\begin{equation}
\frac{\Sigma(N_f=2)}{\Sigma(N_f=3)} = 1.51(11) \text{ in Ref. \cite{Bernard:2012fw}. }
\end{equation}

Regarding the NLO LEC for the decay constant, we get from Fit 1:
\begin{align}
\begin{split}
\frac{L_F(\mu)}{N_c} \cdot 10^3 =&  0.1(4) +  0.6(3) \frac{N_f}{N_c} + O(N_c^{-2}), \label{eq:LFnf}
% \\ \frac{L_M(\mu)}{N_c} \cdot 10^3 =&  -0.2(2) +  2.9(6) \frac{1}{N_c}+ O(N_c^{-2}),  
\end{split}
\end{align}
while for the NLO LEC for the mass, we can only give the $N_c$ scaling at $N_f=4$:
\begin{align}
\begin{split}
 \frac{L_M^{N_f=4}(\mu) }{N_c} \cdot 10^3 
  =&  -0.2(2) +  \frac{2.9(6) }{N_c}+ O(N_c^{-2}).
\end{split}
\end{align}
In the case of Fit 2, we can provide both results together:
\begin{align}
\begin{split}
&\frac{L_F(\mu)}{N_c} \cdot 10^3 =  -0.3(4) +  0.4(4) \frac{N_f}{N_c} + O(N_c^{-2}), \\
&\frac{L_M(\mu)}{N_c} \cdot 10^3 =  -0.1(1) +  0.9(2) \frac{N_f}{N_c}+ O(N_c^{-2}).  \label{eq:LnfNNLO}
\end{split}
\end{align}
From Eqs. \ref{eq:LFnf} and \ref{eq:LnfNNLO}, we can infer the  $N_c=3, N_f=3$ results:
\begin{align}
\begin{split}
{\rm Fit 1}: L_F(\mu) &=  2.1(3) \cdot 10^{-3},   \\% L_M(\mu) &=  1.6(1) \cdot 10^{-3}, \\
{\rm Fit 2}: L_F(\mu) &=  0.4(2.1) \cdot 10^{-3},  \\ L_M(\mu) &=  2.4(8) \cdot 10^{-3}.
\end{split}
\end{align}
%These numbers are  generally compatible with those in Ref. \cite{Aoki:2019cca}. 
%{\bf We would like to stress that standard determinations in QCD yield large error bars in the O(1) NLO LECs of the chiral lagrangian. Our method yields smaller statistical errors, at least concerning the results of Fit 1. }
For $N_c=3, N_f=2$, it is more common to quote ${\bar \ell}_3$ and ${\bar \ell}_4$:
\begin{align}
\begin{split}
\bar{\ell}_3 = 2 \log\left(  {4 \pi F_\pi \over M_\pi^{\text{phys}}}\right) - 16 (4 \pi)^2 L_M^{N_f=2, N_c=3}, \\
\bar{\ell}_4 = 2 \log\left(  {4 \pi F_\pi \over M_\pi^{\text{phys}}}\right) + 4 (4 \pi)^2 L_F^{N_f=2, N_c=3}.
\end{split}
\end{align}
This way, we obtain:
\begin{align}
\begin{split}
\text{Fit 1: }%\bar{\ell}_3 = 2.1(5),  
&\bar{\ell}_4 = 5.1(3), \\
 \text{Fit 2: }&\bar{\ell}_3 = 0.4(1.6), \;\;\;  \bar{\ell}_4 = 4.1(1.1).
\end{split}
\end{align}
We stress that $U(N_f)$ $\bar{\ell}_3$ in fit 2 is not  the same as the standard ${\bar \ell}_3$ in $SU(N_f)$. ${\bar \ell}_4$ agrees instead at $1-2\sigma$ with the results quoted in Ref. \cite{Aoki:2019cca}.

\begin{table*}[th]

\centering

\begin{tabular}{c|c|c|c|c|c|c|c}

Fit & ${F}_0$      & ${F}_1$  & $F_2$  & $(FL_F)^{(0)}$       & $(F L_F)^{(1)}$  & $K_F^{(0)}$   & $\chi^2/dof$ \\ \hline \hline

%1  & 0.0252(2) & -0.039(1)   & 0.64(1) 10^{-3}& 0.0025(5)  & 0.85      \\ \hline

1 & 0.0255(12) & -0.040(6)  & - & $4.7(9.5)\cdot 10^{-6}$ & $4.8(5.1) \cdot 10^{-5}$ &  - & 0.79  \\ \hline

2 & 0.0266(9) & -0.034(8) & -0.033(14) &  $-8(10) \cdot 10^{-6}$ & $5.6(4.4) \cdot 10^{-5}$ & $7.6(6.4)\cdot 10^{-7}$ & 0.9\\ \hline
%2  & 0.0266(8)  & -0.045(6)  & 0.64(1)  10^{-3}   & -0.002(2)  & 1.4   \\ \hline

%3  & 0.0259(7) & -0.040(5)  &  0.64(1) 10^{-3} & -0.001(5) & 1.2\\ \hline

%4  & 0.0255(13) & -0.03(2) &   0.64(1) 10^{-3} & -0.008(8)  & 1.6\\ \hline

\end{tabular}

\caption{Different fits for the decay constant as described in the text.}

\label{tab:allF}

\end{table*}

\begin{table*}[th]

\centering

\begin{tabular}{c|c|c|c|c|c|c|c}

Fit & $B_0$      & $B_1$  & $B_2$ &        $(B L_M)^{(0)}$      & $(B L_M)^{(1)}$    & $K_M^{(0)}$   & $\chi^2/dof$ \\ \hline \hline

1   & 1.70(11) & -0.5(5)    & -  & -0.00046(29)  & 0.0056(15)  & -& 2.0     \\ \hline

%2   & 1.69(6) & -0.41(30)     & -0.00029(12) & 0.0035(6)  & 2.3      \\ \hline

2   & 1.72(7) & -1.8(5)     & 1.8(1.5) & -0.00017(25) & 0.0066(10)  &  $1.3(9) \cdot 10^{-6}$ & 2.4     \\ \hline

%3   & 1.67(15)  & -0.24(84)   &  & 0.0000(4) & 0.0007(24)   & 2     \\ \hline

%4   & 1.79(8)  & -1.38(54)     &  & -0.00026(20) & 0.0041(17)   & 2.3      \\ \hline

\end{tabular}

\caption{Different fits for the meson mass as described in the text.  }

\label{tab:allM}

\end{table*}

\begin{table*}[t]

\centering

\begin{tabular}{c|c|c|c|c}

Fit &  $L_F^{(0)}$       & $L_F^{(1)}$  & $L_M^{(0)}$  & $L_M^{(1)}$ \\ \hline \hline

1 & $1(4)\cdot 10^{-4}$ & $23(13)\cdot 10^{-4}$   & -$20(15)\cdot 10^{-5}$ & $29(6) \cdot 10^{-4}$   \\ \hline

2 & -$3(4)\cdot 10^{-4}$  & $17(18)\cdot 10^{-4}$  & -$1(1)\cdot 10^{-4}$ &  $37(7)\cdot 10^{-4}$  \\ \hline

\end{tabular}

\caption{Values for the LECs from the fits in Tables \ref{tab:allF} and \ref{tab:allM}.}

\label{tab:allLECs}

\end{table*}
%\clearpage

\subsection{Comments on systematics}

The most important systematic uncertainty comes from the finite lattice spacing. Even though a continuum extrapolation would be needed to quantify this error properly, we 
can get an estimate by comparing the pion mass made of different combination of sea and valence quarks.  In particular, Chiral Perturbation Theory in the mixed-action setup predicts that the chiral logs for $F_\pi$ depend upon the mixed pion mass \cite{Chen:2006wf}:
\begin{equation}
M_\pi^{\text{mixed}} \xrightarrow{\text{LO ChPT}} 2B(m^v + m^s),
\end{equation}
where $m^v$ is the renormalized quark mass in the valence sector and $m^s$ in the sea action. We have measured this mixed pion in one ensemble:
\begin{equation}
\text{Ensemble 3A10  } \rightarrow \ aM_\pi^{\text{mixed}} = 0.2201(26),
\end{equation}
obtaining a result which is compatible within errors with both, the sea and valence quark pions.

A different estimate comes from the dependence on $c_{sw}$ in the valence sector. We have recomputed the decay constant  for  $c_{sw} =0$ in the 3A10 ensemble, obtaining 
$[F_\pi]_{c_{sw} =0} = 0.04303(40)$, within $2\%$ of the value at the nominal $c_{sw}$. 
%.  The results are shown in Table \ref{tab:csw0}.
%\begin{align}
%\begin{split}
%\{am^{tm}_0, a\mu\}  = \{-0.9353, 0.0115\}, \ \ c_{sw} = 0
%\end{split}
%\end{align}
%\begin{table}[H]
%\centering
%\begin{tabular}{|c|c|}
%\hline
%$\{am^{tm}_0, a\mu_0\}$ & \{-0.9353, 0.0115\} \\ \hline
%$c_{sw}$                & 0                    \\ \hline
%$aM_\pi$                & 0.2220(19)           \\ \hline
%$am_{pcac}$             & 0.0004(5)            \\ \hline
%$aF_\pi$                & 0.04303(40)          \\ \hline
%\end{tabular}
%\caption{Results and parameters in the valence sector for the ensemble 3A10 with $c_{sw}=0$.}
%\label{tab:csw0}
%\end{table}
%\begin{align}
%\begin{split}
% aM_\pi^{tm} = 0.222&0(19),  am_{pcac} = 0.0004(5),\\ &aF_\pi =0.04303(40).
%\end{split}
%\end{align}
The effects of a change in $c_{sw}$ are in principle $O(a^2)$, which can be estimated at $\sim 2\%$ for this observable. This concerns however only the charged meson sector, since the neutral pion is known to have higher discretization effects with twisted mass. That issue is out of the scope of this work, and it will be addressed in future publications. %We aim to address this topic in further publications.
in $N_c$
{
We end this section with a last word on the chiral fits. We find that our data is well described by ChPT at the order we worked. Still, we cannot exclude that higher order corrections might be relevant in the range of masses we are considering. A robust study on the convergence of ChPT would require simulations at lighter quark masses and a proper continuum extrapolation.
}

\section{Conclusion and Outlook}
\label{sec:conclu}

In this work we presented the first lattice determination using dynamical fermions of the $N_c$ scaling of the couplings in the chiral Lagrangian that contribute to the meson masses and decay constants (see Eqs.~(\ref{eq:fitFpiMev}), (\ref{eq:fitMpiMev}) and Table~\ref{tab:allLECs}). We have been able to disentangle the leading and subleading terms and we found that the subleading contributions are typically non negligible. {In fact, we find that the value for $L_M$ at $N_c=3$ seems to be dominated by the subleading corrections, and the fit result suggests an accidental cancellation of $2L_8 -L_5$ in the large $N_c$ limit.}

% In fact, within statistical precision, the large $N_c$ limit of $L_M$ and $L_F$ is compatible with zero.

From our chiral fits and theoretical expectations, we have been able to infer the values of the couplings for theories with different numbers of flavours, $N_f=2$ and $N_f=3$ at $N_c=3$. We find that our {results nicely} agree with those in the literature regarding $L_F, L_M$ and $F$ (see for example Ref. \cite{Aoki:2019cca} for a summary of results). For $B$ we need to improve our determination, including a non-perturbatively determined renormalization factor. On the other hand, as long as this factor has a small $N_f$ dependence, we can estimate the ratio of B and the chiral condensate for $N_f=2$ and $N_f=3$. We find excellent agreement with the prediction of paramagnetic suppressions of Refs.~\cite{DescotesGenon:1999uh,Bernard:2012fw}. 

\begin{figure*}[tp]
   \centering
   \subfigure[\label{fig:corFpiNLO} Fit 1 for the decay constant.]%
             {\includegraphics[width=0.5\textwidth,clip]{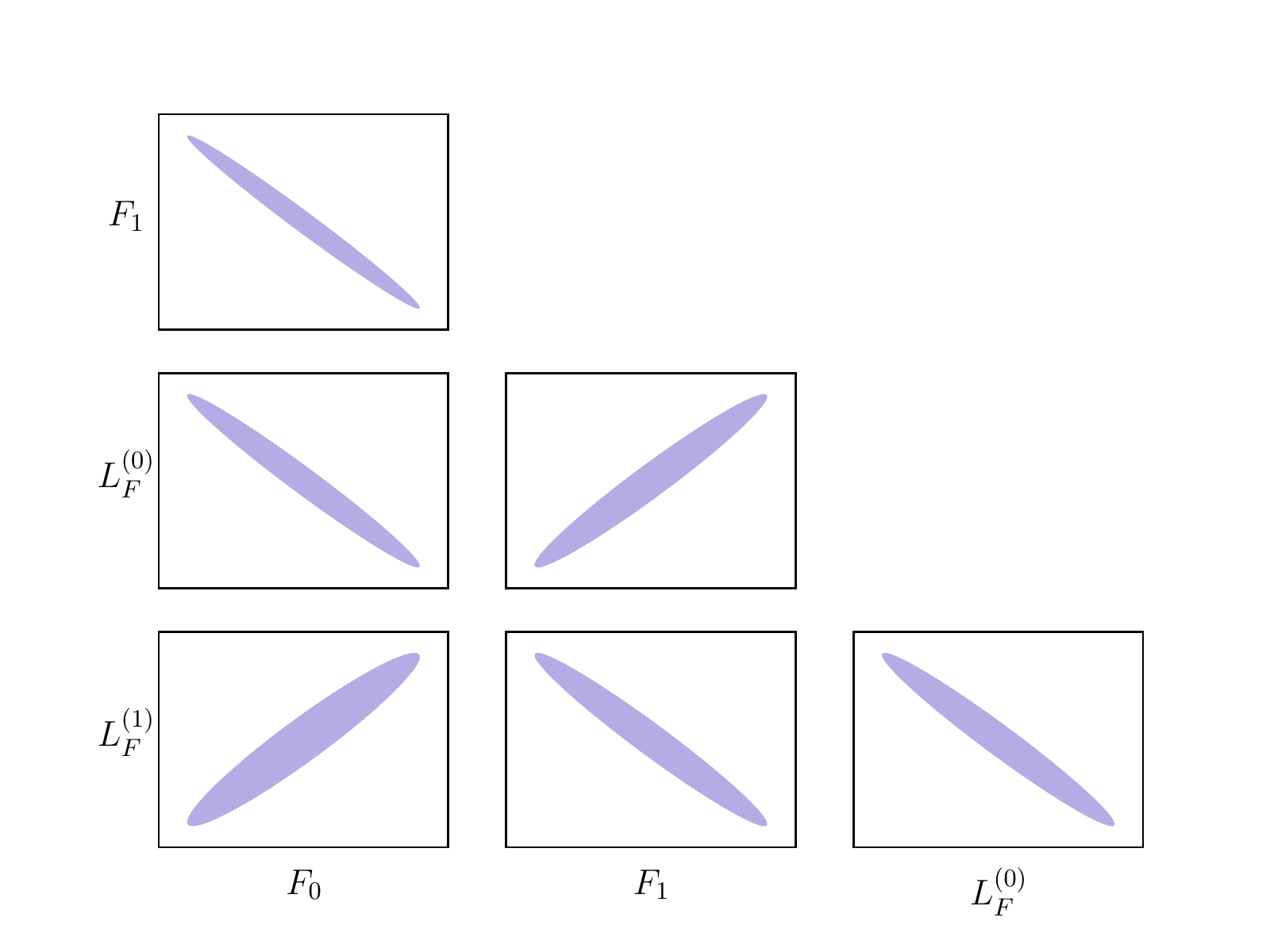}}\hfill
   \subfigure[\label{fig:corMpiNLO} Fit 1 for the mass.]%
             {\includegraphics[width=0.5\textwidth,clip]{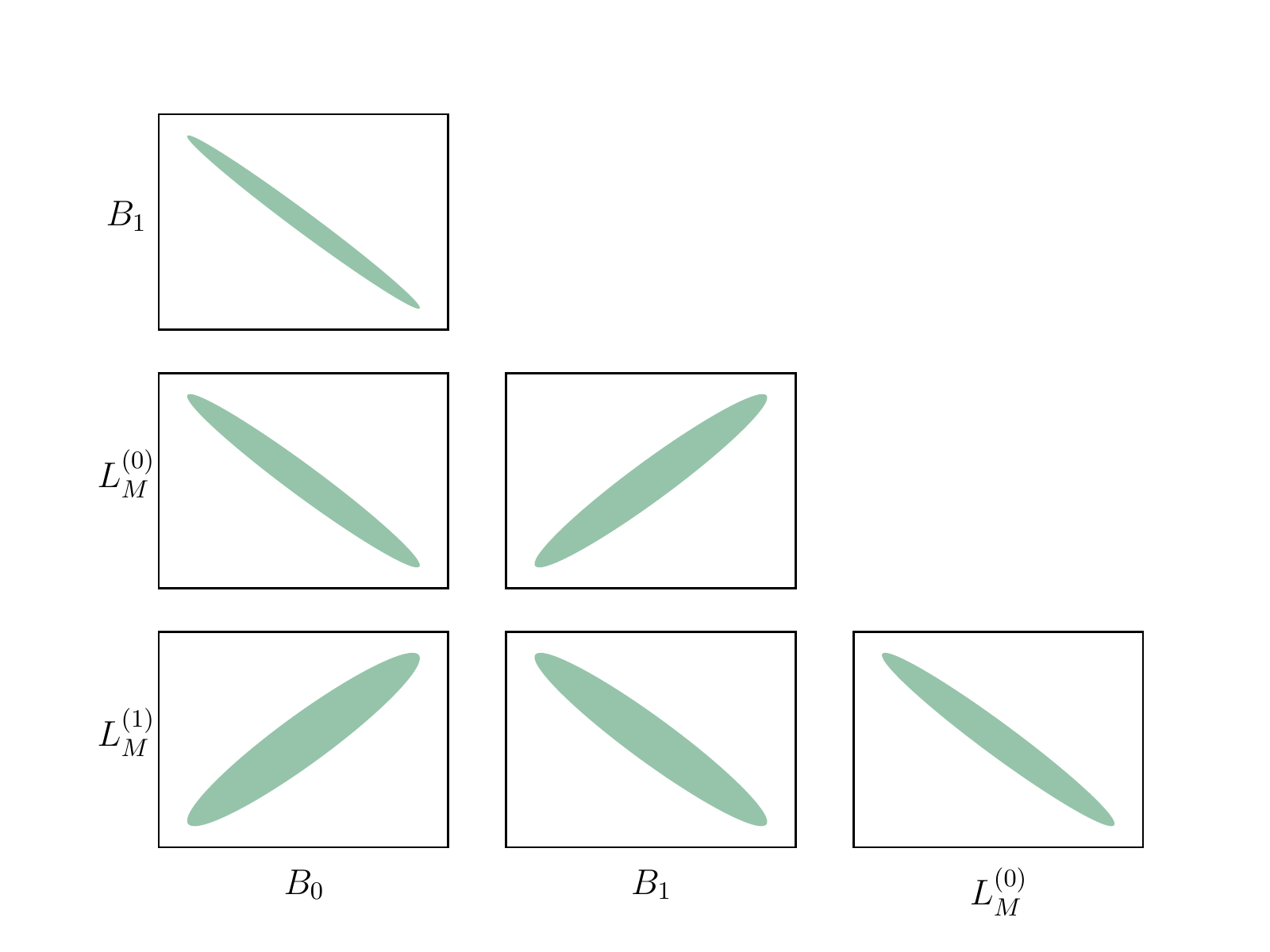}}
   \subfigure[ \label{fig:corFpiNNLO}  Fit 2 for the decay constant.]%
             {\includegraphics[width=0.5\textwidth,clip]{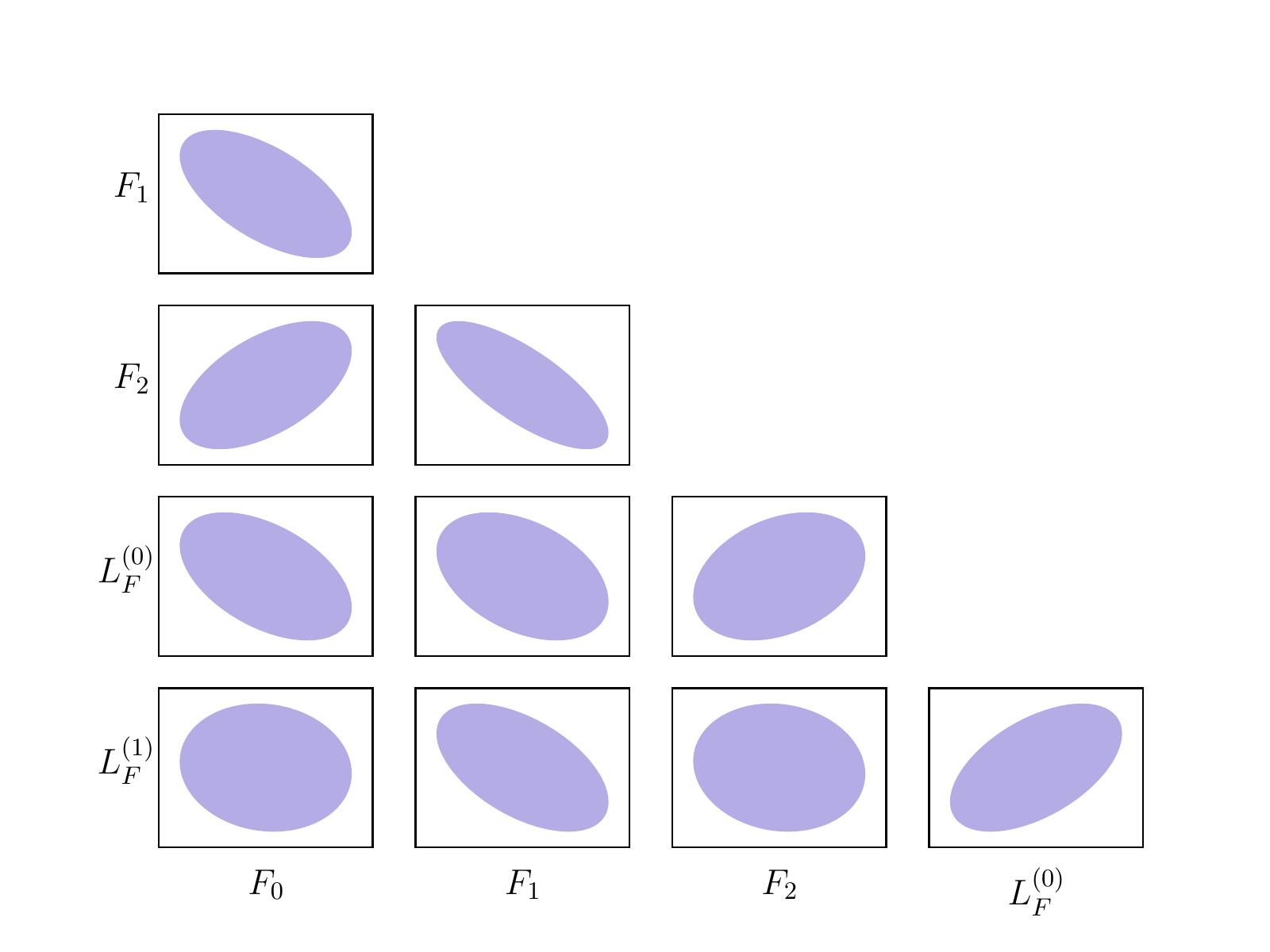}}\hfill
   \subfigure[ \label{fig:corMpiNNLO}  Fit 2 for the mass.]%
             {\includegraphics[width=0.5\textwidth,clip]{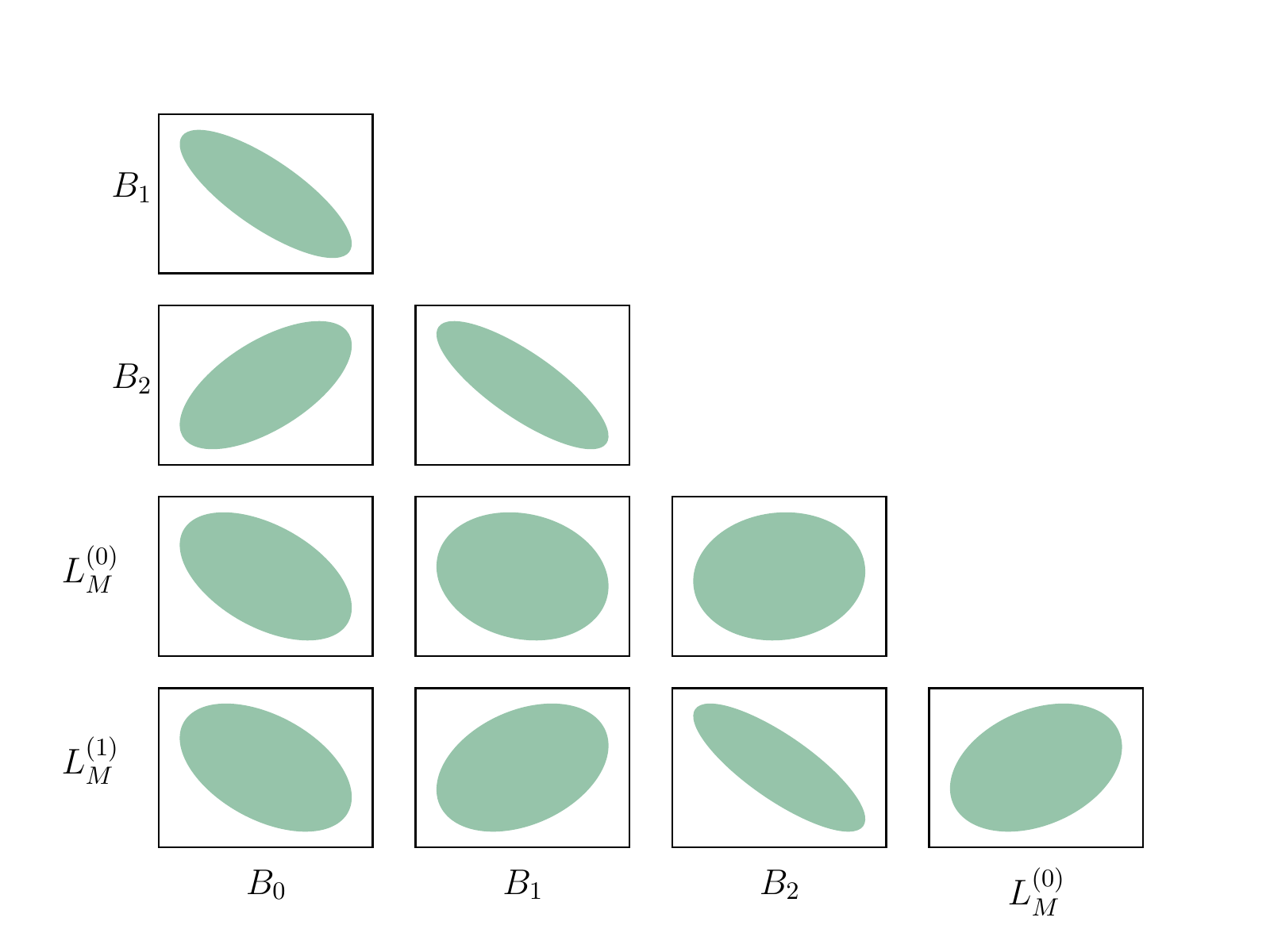}}

   \caption{Correlations between fitted parameters. }
   \label{fig:corr}
\end{figure*}

We would like to stress that the results presented in this paper are complementary to similar studies that can be performed in reduced models \cite{Perez:2018afi, Perez:2016qcv, Gonzalez-Arroyo:2015bya,Hietanen:2009tu,Narayanan:2005gh} or the quenched approximations at large $N_c$ \cite{Bali:2013kia}, since both of these approaches must yield the leading order result as $N_c \to \infty$. Given the strong correlations presents in our results (see Fig. \ref{fig:corr}), a precise determination of the dominant $N_c$ term would significantly improve the determination of the subleading $N_c$ corrections, and hence the determination of the physical values at $N_c=3$. We are willing to provide the bootstrap samples if requested.

As for the future, we would like to mention that our ensembles have a big potential to study other physical observables. We plan to use them to analyse the scaling of other quantities, such as the $K \to \pi$ matrix elements (see \cite{Donini:2016lwz,Romero-Lopez:2018rzy} for previous results). We also believe that the study of scattering amplitudes is a relevant quantity of study at large $N_c$: on one hand quantities such as the $I=2$ $\pi \pi$ scattering length give access to LECs of the chiral Lagrangian; on the other hand the study of the behaviour resonances at large $N_c$ is interesting, as it may shed light about their nature \cite{Pelaez:2010er, Nebreda:2011cp,Bernard:2010fp,RuizdeElvira:2017aet} .

%The former are relevant since the $N_c$ scaling of light scalar resonances is still unclear Pelaez:2010er, Nebreda:2011cp 

%In this work we have been able to compute the leading and subleading contributions of the low energy constants involved in the meson mass and decay constant, $L_M$ and $L_F$. We find that our statistical uncertainties tend to be smaller than those of similar studies at $N_c=3$ (see Ref. \cite{Aoki:2019cca}). In particular, the subdominant LECs are very hard to determine for $N_c=3$ and thus, studying the depences in $N_c$ offers an alternative for better precision.

\vspace{0.5cm}

\begin{acknowledgments}

We thank Andrea Donini for very useful discussions and previous collaboration on related work, as well as  M. Garc\'ia P\'erez, A. Gonz\'alez-Arroyo, G. Herdo\'iza,  A. Ramos, A. Rusetsky, S. Sharpe, C. Urbach and A. Walker-Loud for useful comments and suggestions. 
We are particularly grateful to Claudio~Pica and Martin Hansen for providing us with a $SU(N_c)$ lattice code. This work was partially supported by grant FPA2017-85985-P, MINECO's ``Centro de Excelencia Severo Ochoa'' Programme under grant SEV-2014-0398, 
and the European projects 
H2020-MSCA-ITN-2015/674896-ELUSIVES and H2020-MSCA-RISE-2015/690575-InvisiblesPlus. The work of FRL has also received funding from the European Union Horizon 2020 research and innovation program under the Marie Sk\l{}odowska-Curie grant agreement No. 713673 and "La Caixa" Foundation (ID 100010434, LCF/BQ/IN17/11620044). Furthermore, CP thankfully acknowledges support through the Spanish projects FPA2015-68541-P (MINECO/FEDER) and PGC2018-094857-B-I00, the Centro de Excelencia Severo Ochoa Programme SEV-2016-0597, and the EU H2020-MSCA-ITN-2018-813942 (EuroPLEx)

We thank Mare Nostrum 4 (BSC), Finis Terrae II (CESGA), Tirant 3 (UV) and Lluis Vives (Servei d'Informàtica UV) for the computing time provided.

\end{acknowledgments}

%\vspace{0.5cm}

\bibliography{biblio.bib}

\end{document}